\documentclass[prd,twocolumn,showpacs,floatfix,preprintnumbers,letterpaper]{revtex4}
\usepackage{graphicx}
\usepackage{epsfig}
\usepackage{amsfonts}

\newcommand{\md}{{\rm d}}

\begin{document}

\title{Canonical Gravity with Fermions}
\author{Martin Bojowald}
\affiliation{Institute for Gravitation  and the Cosmos, The
  Pennsylvania State University,
104 Davey Lab, University Park, PA 16802}\author{Rupam Das} 
\affiliation{Department of Physics and Astronomy, Vanderbilt University,
Nashville, TN 37235}
\affiliation{Institute for Gravitation and the Cosmos, 
The Pennsylvania State University,
104 Davey Lab, University Park, PA 16802}

\begin{abstract}
  Canonical gravity in real Ashtekar--Barbero variables is generalized
  to allow for fermionic matter. The resulting torsion changes several
  expressions in Holst's original vacuum analysis, which are
  summarized here. This in turn requires adaptations to the known loop
  quantization of gravity coupled to fermions, which is discussed on
  the basis of the classical analysis. As a result, parity invariance
  is not manifestly realized in loop quantum gravity.
\end{abstract}
\maketitle
\section{INTRODUCTION}
\label{sec:INTRODUCTION}

When matter is considered coupled to classical or quantum gravity,
several important issues arise for fermions. This is, e.g., related to
the chirality and possible parity violation of spinors or the fact
that they contribute torsion to the space-time geometry. In loop
quantum gravity, fermions have been treated occasionally but not yet,
as detailed below, in a complete manner. They are therefore revisited
here especially with canonical quantization in mind.

The canonical formulation of general relativity in complex Ashtekar
variables \cite{AshVar} recasts gravity as a gauge theory similar to
Yang-Mills theory, which offered a new way to a possible quantum
theory of gravity. Although this reformulation of gravity, expressed
in Ashtekar variables as a dynamical theory of complex-valued
connections, has the advantage of obtaining algebraically simple
constraints, rather complicated reality conditions have to be imposed
on the basic canonical variables in order to recover real, Lorentzian
general relativity. Moreover, since holonomies of the complex Ashtekar
connections take values in the non-compact gauge group ${\rm
  SL}(2,{\mathbb C})$, this approach prevents one from taking
advantage of much of the available mathematical arsenal of gauge
theory built upon compact gauge groups.  Therefore, real su(2) valued
Ashtekar-Barbero connections, that is,
$A^{i}_{a}=\Gamma^{i}_{a}+\gamma K^{i}_{a}$ (with the spin connection
$\Gamma^{i}_{a}$, $K^{i}_{a}$ being a 1-form derived from extrinsic
curvature and the Barbero--Immirzi parameter $\gamma$
\cite{AshVarReell,Immirzi} taking any non-zero real value), have
mainly been used for the passage to a quantum theory of gravity.

Real variables were initially introduced by Barbero in a purely
canonical formalism \cite{AshVarReell} which left the relation of the
real connection to possible pull-backs or projections of space-time
objects unclear.  Holst, motivated by this issue, carried out an
analysis in \cite{HolstAction} to re-derive Barbero's canonical
formulation from an action which generalizes the ordinary
Hilbert-Palatini action. In this paper, we further generalize Holst's
analysis for pure gravity to allow for fermionic matter. In other
words, we present the Hamiltonian formulation of the Einstein--Cartan
action, which incorporates Holst's action for the gravitational part.
This issue has been considered in the literature several times, but
the available discussions appear incomplete.  In addition to filling
this gap in the classical analysis, details of the canonical
formulation whose results we summarize are crucial for a proper
quantization of gravity in the presence of fermions.

In particular, non-zero torsion arising from the coupling of fermionic
matter to gravity through the spin connection requires an analysis in
terms of more general connections than used in Holst's analysis, which
inherit torsion contributions. Our results for the given
Einstein--Cartan action, despite some resemblance to those in
\cite{SugraAshtekar,QSDI,FermionAshtekar}, differ in several
details. Moreover, we generalize the canonical treatment to arbitrary
non-minimal coupling of fermions without any inconsistencies as they
occur in other approaches.

We summarize those derivations in a classical part in this paper,
which are important to see the role of parity.  These details will
show us the crucial changes implied by torsion for the general form of
dynamics as well as parity invariance, and thus also play a role for
any quantization based on a formulation in Ashtekar variables.
Consequences for a loop quantization and its dynamics, where several
ingredients depend on the form of connections and the phase space
structure, are thus described in the second and main part of this
paper.  Necessary adaptations to the loop quantization of gravity with
fermions are explored and presented with the conclusion that previous
constructions go through but require non-trivial changes. In
particular, the form of basic variables used in the quantum
representation makes it difficult to prove parity invariance of the
quantum theory even if no parity violating classical interactions are
used. This leaves open the potentially intriguing possibility that
loop quantum gravity may provide small parity breaking effects due to
the quantum space-time structure in the presence of fermions and
torsion.

\section{Canonical Formulation}
\label{sec:canonical Formulation}

For fermions, one can use a tetrad $e^I_{\mu}$ rather than a
space-time metric $g_{\mu\nu}$, related by
$e^I_{\mu}e^I_{\nu}=g_{\mu\nu}$, in order to formulate an action with
the appropriate covariant derivative of fermions. This naturally leads
one to a first-order formalism of gravity in which the basic
configuration variables are a connection 1-form and the tetrad. In
vacuum the connection would, as a consequence of field equations, be
the torsion-free connection compatible with the tetrad.  In the
presence of matter fields which couple directly to the connection,
such as fermions, this is no longer the case and there is torsion
\cite{SpinTorsion}.  For completeness and to introduce the notation,
we start by demonstrating this well-known origin of torsion in the
theory of gravity non-minimally coupled to fermionic matter.

\subsection{Einstein--Cartan Action}
\label{sec:Einstein-Cartan Action}

The basic configuration variables in a Lagrangian formulation of
fermionic field theory are the Dirac bi-spinor $\Psi =
\left(\psi,\eta\right)^T$ and its complex conjugate in
$\overline{\Psi}=\left(\Psi^{*}\right)^{T}\gamma^{0}$ with
$\gamma^{\alpha}$ being the Minkowski signature Dirac matrices. We
note that $\psi$ and $\eta$ transform with density weight zero and are
spinors according to the fundamental representations of ${\rm
SL}\left(2,{\mathbb C} \right)$.  Then the non-minimal coupling of
gravity to fermions can be expressed by the total action composed of
the gravitational contribution $S_G$ and the matter contribution $S_F$
resulting from the fermion field:
\begin{widetext}
\begin{eqnarray}
\label{nonminimalaction}
S\left[e,\omega,\Psi\right]&=& S_{G}\left[e,\omega\right]+
S_{F}\left[e,\omega,\Psi\right]  \\ &=& 
\frac{1}{16\pi G} \int_{M}\md^{4}x \;|e|e^{\mu}_{I}e^{\nu}_{J}
P^{IJ}_{\ \ \ KL}F^{\ \ KL}_{\mu
\nu}(\omega) + \frac{1}{2}i\int_{M} \md^{4}x \;
|e|\left[\overline{\Psi}\gamma^{I}e^{\mu}_{I}\left(1-
\frac{i}{\alpha}\gamma_{5}\right)\nabla_{\mu}\Psi -
\overline{\nabla_{\mu}\Psi}\left(1-\frac{i}{\alpha}\gamma_{5}\right)
\gamma^{I}e^{\mu}_{I}\Psi\right],\nonumber
\end{eqnarray}
\end{widetext}
where $\alpha\in{\mathbb R}$ is the parameter for non-minimal
coupling. The $\alpha$-dependent terms of this form have been
introduced in \cite{FermionAshtekar} to generalize results of
\cite{FermionImmirzi,FermionTorsion} (see also
\cite{FermionImmirziNonMin}), where they played important (though
indirect) roles in parity properties \footnote{In the gravitational
part, however, we will not follow exactly the notation of
\cite{FermionAshtekar} but rather that of \cite{HolstAction}}.  Here
$I,J,\ldots=0,1,2,3$ denote the internal Lorentz indices and $\mu,
\nu, \ldots=0,1,2,3$ the respective space-time indices.  For
simplicity, we ignore fermionic mass terms or potentials as they do
not provide further complications.

The first term in (\ref{nonminimalaction}) is the Holst action
\cite{HolstAction} of gravity \footnote{At this point, it is
noteworthy that we intend to use the signature $(- + + +)$ (instead of
$(+ - - -)$ which is common in quantum field theory) for both gravity
and fermions since this is the signature most prevalent in the
literature for canonical gravity. This demands certain modifications
in the representations of the Clifford algebra, where it turns out
that changing the signature from $(+ - - -)$ to $(- + + +)$ only
requires all the Dirac matrices to be multiplied by $i$.},
$e^{\mu}_{I}$ is the tetrad field, $e$
is its determinant, and $e^{I}_{\mu}$ its inverse. The Lorentz
connection in this formulation is denoted by $\omega_{\mu}^{IJ}$ and
$F^{KL}_{\mu \nu}(\omega)= 2\partial_{[\mu}\omega^{IJ}_{\nu]}+
\left[\omega_{\mu},\omega_{\nu}\right]^{IJ}$ is its curvature. In
order to write the Holst action in a compact form, we have used the
following tensor and its inverse
\begin{eqnarray}
\label{PIJ}
P^{IJ}_{\ \ \ KL}&=&\delta^{[I}_{K} \delta^{J]}_{L} - \frac{1}{\gamma}
\frac{\epsilon^{IJ}_{\ \ KL}}{2}\\
{P^{-1}_{\ \ \
IJ}}^{KL}&=&\frac{\gamma^{2}}{\gamma^{2}+1}\left(\delta^{[K}_{I}
\delta^{L]}_{J} + \frac{1}{\gamma} \frac{\epsilon_{IJ}^{\ \ \
KL}}{2}\right)\nonumber
\end{eqnarray}
where $\gamma$ is again the Barbero--Immirzi parameter.  Finally, the
covariant derivative $\nabla_{\mu}$ of Dirac spinors is defined by
\begin{equation}
\label{covariantderivative}
 \nabla_{\mu}\equiv \partial_{\mu} + \frac{1}{4}\omega^{IJ}_{\mu}
 \gamma_{[I} \gamma_{J]}\quad, \ \ \ \ \
 \left[\nabla_{\mu},\nabla_{\nu}\right] = \frac{1}{4}F^{IJ}_{\mu
 \nu}\gamma_{[I} \gamma_{J]}
\end{equation}
in terms of Dirac matrices $\gamma_I$ (which will always carry an
index such that no confusion with the Barbero--Immirzi parameter should
arise).  Note that we are ignoring the gauge connection required for
describing an interaction between charged fermions in the definition
of the covariant derivative (\ref{covariantderivative}). However, this
analysis can easily be generalized to incorporate such interactions.

Varying the action by $\omega_{\mu}^{IJ}$ produces equations which can
be solved for the torsion contribution in $\left(\nabla_{\mu}-
\widetilde{\nabla}_{\mu}\right)V_{I} = C_{\mu I}^{\ \ \ J} V_{J}$
where $\widetilde{\nabla}_{\mu}$ is the covariant derivative
compatible with the tetrad:
\begin{eqnarray}
\label{CIJK}
e^{\mu}_{I}C_{\mu JK}
&=&2\pi G
\frac{\gamma}{\gamma^{2}+1} \left(\beta \epsilon_{IJKL}J^{L} - 
     2{\theta} \eta_{I[J} J_{K]}\right)
\end{eqnarray}
where $\beta:=\gamma+1/\alpha$, $\theta:=1-\gamma/\alpha$.  This
contorsion tensor depends on the Immirzi parameter $\gamma$ unless
$\alpha=\gamma$. This can then be inserted to produce an action of the
form
\begin{widetext}
\begin{eqnarray}
\label{interactingaction1}
S\left[e,\omega,\Psi\right]&=& S_{G}\left[e,\widetilde{\omega}\right]+
S_{F}\left[e,\widetilde{\omega},\Psi\right]+
S_{\rm int}\left[e,\Psi\right] \nonumber\ \\&=& \frac{1}{2\kappa}
\int_{M}\md^{4}x \;|e|e^{\mu}_{I}e^{\nu}_{J}\widetilde{F}^{\ \ IJ}_{\mu \nu}(\widetilde{\omega}) +
\frac{1}{2}i\int_{M} \md^{4}x \;
|e|\left(\overline{\Psi}\gamma^{I}e^{\mu}_{I}
{\widetilde{\nabla}}_{\mu}\Psi
- \overline{{\widetilde{\nabla}}_{\mu}\Psi}
\gamma^{I}e^{\mu}_{I}\Psi\right)
\nonumber
\\ &&+\frac{3\kappa}{16} \frac{\gamma^{2}}{\gamma^{2}+1}
\left(\frac{1}{\alpha^{2}}-\frac{2}{\alpha \gamma}-1\right)\int_{M}\md^{4}x
\;|e|
(\overline{\Psi}\gamma_{5}\gamma_{L}\Psi)(\overline{\Psi}
\gamma^{5}\gamma^{L}\Psi), 
\end{eqnarray}
\end{widetext}
with a simple interaction term in addition to gravity and fermion
contributions of the torsion-free form.

Notice that the second term in the gravitational Holst action
containing $\gamma$ and the term involving non-minimal coupling
$\alpha$ in the Dirac action are dropped from the above effective
action since both these terms can be expressed as boundary terms {\em
on-shell}; see \cite{HolstAction} for details concerning the second
term in Holst action. The non-minimally coupled term in the Dirac
action can be cast as a boundary term after using
$\tilde{\nabla}_{\mu}(ee^{\mu}_N)=0$ on solutions:
\begin{widetext}
\begin{eqnarray}
\label{nmterm}
\frac{1}{2\alpha}\int_{M} \md^{4}x \;
|e|\left(\overline{\Psi}\gamma^{I}e^{\mu}_{I}
\gamma_{5}{\widetilde{\nabla}}_{\mu}\Psi
-\overline{{\widetilde{\nabla}}_{\mu}\Psi}
\gamma_{5}\gamma^{I}e^{\mu}_{I}\Psi\right)
&=& \frac{1}{2\alpha}\int_{M} \md^{4}x \;
|e| e^{\mu}_{I}\left({\partial}_{\mu}({\overline{\Psi}}
\gamma^{I}\gamma_{5}\Psi)-\frac{1}{4}{\widetilde{\omega}}_{\mu}^{MN}
{\overline{\Psi}}\gamma_{5}\left[\gamma^{I},\gamma_{[M}\gamma_{N]}
\right]_{-}\Psi\right)\\
&=& \frac{1}{2\alpha}\int_{M} \md^{4}x \;
|e| \left(e^{\mu}_{I}({\partial}_{\mu}J^{I})+{\widetilde{\omega}}_{\mu}^{MN}
e^{\mu}_{M}J_{N}\right)=
\frac{1}{2\alpha}\int_{M} \md^{4}x \;
 {\partial}_{\mu}(|e|e^{\mu}_{I}J^{I}) \nonumber
\end{eqnarray}
\end{widetext}
with the axial current $J^I= \overline{\Psi}\gamma^I\gamma_5\Psi$. In
particular, as we will discuss in more detail in Sec.~\ref{s:Parity},
the effective action (\ref{interactingaction1}) is parity invariant
{\em for all real} $\alpha$. However, as noted in
\cite{FermionAshtekar}, there is an indirect effect of parity because
not all torsion components transform as expected under parity unless
$\alpha=\gamma$. We will also see this in the canonical description in
what follows, before discussing its significance in classical and
quantum gravity. Moreover, for $\alpha=\gamma$ the action becomes
completely independent of $\gamma$ as noted in
\cite{FermionAshtekar}. In this case, we have equations for fermions
minimally coupled to Einstein--Cartan gravity rather than gravity
described by the Holst action. This case is also geometrically
distinguished by topological invariance properties of boundary terms.

\subsection{Canonical variables and second class constraints}

To set up a Hamiltonian formalism, one foliates space-time into
spatial slices $\Sigma_t\colon t={\rm const}$ determined by a time
function $t$. Instead of working with space-time tensors, one uses
spatial tensors which depend on $t$, subject to evolution equations
along a time evolution vector field $t^{\mu}$ such that
$t^{\mu}\nabla_{\mu}t=1$. Since we are using the Lorentzian signature,
the vector field $t^{\mu}$ is required to be future directed.  Let us
decompose $t^{\mu}$ into normal and tangential parts with respect to
$\Sigma_{t}$ by defining the lapse function $N$ and the shift vector
$N^{a}$ as $t^{\mu}= Nn^{\mu}+ N^{\mu}$ with $N^{\mu}n_{\mu}=0$, where
$n^{\mu}$ is the future directed unit normal vector field to the
hypersurfaces $\Sigma_{t}$.  The space-time metric $g_{\mu\nu}$
induces a spatial metric $q_{\mu\nu}$ by the formula
$g_{\mu\nu}=q_{\mu\nu}-n_{\mu}n_{\nu}$.  Since contractions of
$q_{\mu\nu}$ and $N^{\mu}$ with the normal $n^{\mu}$ vanish, they give
rise to spatial tensors $q_{ab}$ and $N^a$. Here, the lower case roman
letters, $a, b, c, \ldots$ , are used to imply spatial tensorial
indices.

Moreover, since we are using a tetrad formulation, in addition to the
above foliation of the space-time manifold we perform a partial gauge
fixing on the internal vector fields of the tetrad to decompose it
into an internal unit time-like vector and a triad. Let us fix a
constant internal vector field $n_{I}=-\delta_{I,0}$ with
$n^{I}n_{I}=-1$. Now we allow only those tetrads which are compatible
with the fixed $n^{I}$ in the sense that $n^{a}:= n^{I}e_{I}^{a}$ must
be the unit normal to the given foliation. This implies that
$e^{a}_{I}= {\cal E}^{a}_{I} - n^{a}n_{I}$ with ${\cal
E}^{a}_{I}n_{a}={\cal E}^{a}_{I}n^{I}=0$ so that ${\cal E}^{a}_{I}$ is
a triad.
 
Now using $n^{a}=N^{-1}(t^{a}-N^{a})$ to project fields normal and
tangential to $\Sigma_{t}$, one can decompose the Einstein--Cartan
action (\ref{nonminimalaction}) and extract the canonical fields as
well as possible constraints. The only time derivative (along
$t^{\mu}$) of gravitational variables in the action appears for the
Ashtekar--Barbero connection
\begin{equation}
\label{+a}
A^{j}_{b}:=-\gamma\omega_{b}^{\
  j0}-\frac{1}{2}\epsilon^{j}_{\  kl}\omega_{b}^{\  kl}= 
\gamma K_b^j+\Gamma_b^j 
\end{equation}
multiplied with $P^a_i:=\sqrt{q}{\cal E}^a_i/\gamma\kappa$ as the
momentum conjugate to $A_b^j$. The remaining components of the
space-time connection,
\begin{equation}
{\,^{-}\!\!A}^{j}_{b}:=\omega_{b}^{\
  j0}-\frac{1}{2\gamma}\epsilon^{j}_{\  kl}\omega_{b}^{\  kl} 
\end{equation}
and $\omega_t^{IJ}$, as well as the lapse function $N$ and shift
vector $N^a$ appearing in the metric are non-dynamical. As usually,
variation by $N$ and $N^a$ gives the Hamiltonian and diffeomorphism
constraints. The variation by the non-dynamical connection components,
on the other hand, provides partially second class constraints which
can be solved algebraically for
\begin{eqnarray}
\label{gammabk2}
\gamma {\,^{-}\!\!A}_{b}^{k}&=&-A_{b}^{k}+
 2{\Gamma}^{k}_{b}
\end{eqnarray}
where
\begin{eqnarray}
\label{gammabk1}
\Gamma^{k}_{b}&=&
\widetilde{\Gamma}^{k}_{b}+\frac{\gamma\kappa}{4(1+\gamma^{2})}
\left(\theta 
\ \epsilon_{ij}^{\ \  k}e_{b}^{i}J^{j}-\beta e_{b}^{k} J^{0}\right)\,, 
\end{eqnarray} 
is a combination of the torsion-free spin connection
$\widetilde{\Gamma}^k_b$ (see
App.~\ref{appen:SpinconnectionTorsion}) and a torsion contribution
\begin{eqnarray}
\label{cbk1}
C^{j}_{a} 
&:=& \frac{\gamma
  \kappa}{4(1+\gamma^{2})}\left(\theta \ \epsilon^{j}_{\
    kl}e_{a}^{k}J^{l}- \beta e_{a}^{j} J^{0}\right)\,.
\end{eqnarray}
Also $\omega_t^{k0}$ is determined by the second class constraints,
such that only $\epsilon_{ijk}\omega_t^{jk}$ remain free as Lagrange
multipliers of the Gauss constraint
\begin{equation}
\label{gc1}
G_i = {\cal D}_{b}P^{b}_{i}-\frac{1}{2}\sqrt{q}J_{i}
 = \gamma [K_{b},P^{b}]_{i} -
\frac{\gamma \beta}{2(1+\gamma^{2})}\sqrt{q}J_{i}
\end{equation}
where $J^i =\psi^{\dagger}\sigma^{i}\psi+\eta^{\dagger}\sigma^{i}\eta$
(and $J^{0} = \psi^{\dagger}\psi-\eta^{\dagger}\eta$ which will appear
below).  

Together with the diffeomorphism constraint
\begin{widetext}
\begin{eqnarray}
\label{totaldiffconstraint}
C_a&=& P^{b}_{j}(F_{ab}^{j}-(\gamma^{2}+1)
\epsilon^{j}_{\ kl}K^{k}_{a}K^{l}_{b})
-i \sqrt{q}\left(\theta_{L}(\psi^{\dagger}{\cal D}_{a}{\psi}-
\overline{{\cal D}_{a}{\eta}}\eta)- c.c.\right)+
\frac{\beta}{2}K_{a}^{i}\sqrt{q}J_{i}\\
&=&
 P^{b}_{j}F_{ab}^{j}-i \sqrt{q}\left(\theta_{L}(\psi^{\dagger}
{\cal D}_{a}{\psi}-
\overline{{\cal D}_{a}{\eta}}\eta)- c.c.\right)-
\frac{\gamma^{2}+1}{\gamma}K^{j}_{a}G_{j} \label{newtotaldiffconstraint}
\end{eqnarray}
and the Hamiltonian constraint
\begin{eqnarray}
\label{totalhamiltonian}
C \ &=& 
\frac{\gamma^{2}\kappa}{2\sqrt{q}}{P}^{a}_{i}P^{b}_{j}\left(\epsilon^{ij}_{\
  \ k}F_{ab}^{k}-2(\gamma^{2}+1)K_{[a}^{i}K_{b]}^{ j}\right)+\frac{\gamma
  \kappa \beta}
      {2\sqrt{q}}{P}_{i}^{a}{\cal D}_{a}(\sqrt{q}J^{i})+(1+\gamma^{2})
\kappa\widetilde{D}_{a}
\left(\frac{{P}_{i}^{a}G^{i}}{\sqrt{q}}\right)\nonumber\ \\ 
&&+i\gamma\kappa
P^{a}_{i}\left(\theta_{L}(\psi^{\dagger}\sigma^{i}{\cal
  D}_{a}\psi+\overline{{\cal D}_{a}\eta}\sigma^{i}\eta))-
\theta_{R}(\eta^{\dagger}\sigma^{i}{\cal
  D}_{a}\eta+\overline{{\cal D}_{a}\psi}\sigma^{i}\psi)\right) 
+\frac{\kappa}{4}\left(3-\frac{\gamma}{\alpha}+
2\gamma^{2}\right)\epsilon_{lkn} K_{a}^{l}P^{a}_{k}J^{n}
\end{eqnarray}
\end{widetext}
(where $\theta_{L/R}:= \frac{1}{2}(1\pm i/\alpha)$) this provides a first
class set of constraints.

At this point, we emphasize that we have not imposed any restriction
on either the non-minimal coupling parameter $\alpha$ or the Immirzi
parameter $\gamma$ (as long as they are both real). The formulation is
thus consistent for all values, but as we will see the behavior under
parity of the variables used appears different depending on whether
$\alpha=\gamma$ or not. We also emphasize that some of the terms in
our constraints differ from those presented in \cite{FermionAshtekar}
even for the case $\alpha=\gamma$ considered there. In what follows,
we will be led to consistency checks of our expressions, which confirm
the presence of the terms listed here.

\section{Parity transformation of the classical theory}
\label{s:Parity}

In the presence of fermions, the parity behavior is not fully obvious
even in the absence of explicitly parity violating interaction
terms. A detailed analysis of transformation properties is then
required.

\subsection{The Torsion Contribution to Extrinsic Curvature}
\label{TorsionK}

Torsion components play an indirect but important role in the behavior
under parity.  During the constraint analysis, second class
constraints provide the torsion contribution to the connection as seen
in (\ref{gammabk1}). However, although $K^{i}_{a}$ is restricted by
the Gauss constraint, constraints do not provide its complete torsion
contribution.  On the other hand, the transformation properties of the
Ashtekar-Barbero connection $A_a^i$ under parity cannot be determined
without the knowledge of the torsion contribution to $K_a^i$, or at
least its parity behavior. Thus the splitting of extrinsic curvature
into torsion-free and torsion parts is inevitable in order to arrive
at a set of consistent parity transformations for gravity with
fermions.  As in the case of (\ref{CIJK}), we have to solve partially
equations of motion for the connection to derive the expression for
the torsion part $k^{i}_{a}$ of $K^{i}_{a}= \tilde{K}_a^i+k_a^i$.

For the canonical pair $(A_{a}^{i},P^{a}_{i})$ the equations of motion
are ${\cal L}_{t}A_{a}^{i} = \left\{A_{a}^{i}, H\right\} = \delta
H/\delta P^{a}_{i}$ and ${\cal L}_{t}P^{a}_{i} = \left\{P^{a}_{i}, H
\right\} = -\delta H/\delta A_{a}^{i}$ where
$H[\Lambda^i,N,N^a]=\int\md^3x (\Lambda^iG_i+NC+N^aC_a)$ is the total
constraint.  While the first equation of motion entails all the
dynamics of gravity coupled with matter, the second one yields the
expression for the connection. After longer calculations, it takes the
form
\begin{eqnarray*}
&&{\cal L}_{t}P^{c}_{j}+\omega^{i0}_t\epsilon_{ij}{}^k
  P^{c}_{k}-P^{c}_{j}\partial_{a}N^{a}-
N^{a}\partial_{a}P^{c}_{j}+P^{b}_{j}\partial_{b}N^{c}\nonumber\\
&&+ N^{a}\epsilon^{i}_{\ jk}P^{c}_{i}A^{k}_{a}+N^{c}G_{j}
+{\rm sgn}\det (e_a^i)\frac{\epsilon^{abc}}{\gamma^{2}\kappa}\partial_{b}
(Ne_{aj})
\\
&&+\frac{N\sqrt{q}}{\gamma^{2}\kappa}(e^{b}_{j}e^{c}_{k}-
e^{b}_{k}e^{c}_{j})A_{b}^{k} 
\ = \ \frac{1}{2}\kappa N\epsilon^{i}_{\ jk}P_{i}^{c}J^{k}+
\frac{N \kappa}{2\alpha}P_{j}^{c}J^{0}\nonumber
\end{eqnarray*}
where the sign of the determinant of the co-triad appears due to
the use of $\gamma\kappa P^a_i=\frac{1}{2}{\rm sgn}\det (e_a^i)
\epsilon^{abc}\epsilon_{ijk} e^j_be^k_c$.

In order to solve for $k_{a}^{i}$ which appears via $A_a^i$, we
contract the equation with $e_{c}^{l}$ and, as an internal tensor with
indices $l$ and $j$, derive its trace and symmetric parts. Combined,
this gives
\begin{widetext}
\begin{eqnarray}
\label{eom3}
e_{c}^{l}{\cal L}_{t}P^{c}_{j}&+&e_{c}^{j}{\cal
  L}_{t}P^{c}_{l}-\delta^{l}_{j}e_{c}^{k}{\cal 
  L}_{t}P^{c}_{k}+N^{a}(P^{c}_{j}\partial_{a}e^{l}_{c}+
  P^{c}_{l}\partial_{a}e^{j}_{c})+e_{c}^{l}P^{b}_{j}\partial_{b}N^{c}+
  e_{c}^{j}P^{b}_{l}\partial_{b}N^{c}\\ &-&
  {\rm sgn}\det(e_a^i)\left(\delta^{l}_{j} 
\frac{\epsilon^{abc}}{\gamma^{2}\kappa} \
  Ne_{c}^{l}\partial_{b}(e_{aj})- N
  \frac{\epsilon^{abc}}{\gamma^{2}\kappa}
  (e_{c}^{l}\partial_{b}e_{aj}+ e_{c}^{j}\partial_{b}e_{al})\right)
  +\frac{N\sqrt{q}}{\gamma^{2}\kappa}
  (e^{b}_{j}A_{b}^{l}+e^{b}_{l}A_{b}^{j}) \ = \ -\frac{N\sqrt{q}}{2\alpha}
\delta^{l}_{j}J^{0}\,.  \nonumber
\end{eqnarray} 
The extrinsic curvature contribution is contained in the term
\begin{eqnarray}
\label{aexpansion}
\frac{\sqrt{q}}{\gamma\kappa}(e^{b}_{j}A_{b}^{l}+e^{b}_{l}A_{b}^{j})
\ = \
(P^{b}_{j}\widetilde{\Gamma}_{b}^{l}+P^{b}_{l}
\widetilde{\Gamma}_{b}^{j})+
(P^{b}_{j}C_{b}^{l}+P^{b}_{l}C_{b}^{j})+ 
\gamma(P^{b}_{j}\widetilde{K}_{b}^{l}+P^{b}_{l}\widetilde{K}_{b}^{j})+
\gamma(P^{b}_{j}k_{b}^{l}+P^{b}_{l}k_{b}^{j})\,, 
\end{eqnarray}
where we have used the decomposition
$A_{a}^{i}={\widetilde{A}}^{i}_{a}+{\overline{A}}^{i}_{a}$ into the
torsion-free part
$\widetilde{A}_{a}^{i}=\widetilde{\Gamma}^{i}_{a}+\gamma
\widetilde{K}^{i}_{a}$ and a torsion contribution
$\overline{A}_{a}^{i}=C^{i}_{a}+\gamma k^{i}_{a}$. 

To complete the splitting, we use that the torsion-free extrinsic
curvature from the usual expression $\tilde{K}_{ab} =
\frac{1}{2N}(\dot{q}_{ab}-2\widetilde{D}_{(a}N_{b)})$ satisfies
\begin{eqnarray}
\label{k1}
P^{b}_{j}\tilde{K}_{b}^{l}+P^{b}_{l}\tilde{K}_{b}^{j}
&=& -\frac{1}{N}{\gamma}\left((e_{c}^{l}{\cal L}_{t}P^{c}_{j}+
e_{c}^{j}{\cal L}_{t}P^{c}_{l}-\delta^{l}_{j}e_{c}^{k}{\cal
  L}_{t}P^{c}_{k}) +
(N^{a}(P^{c}_{j}\partial_{a}e^{l}_{c}+P^{c}_{l}\partial_{a}e^{j}_{c})+
e_{c}^{l}P^{b}_{j}\partial_{b}N^{c}+e_{c}^{j}P^{b}_{l}
\partial_{b}N^{c})\right)
\end{eqnarray} 
\end{widetext}
for $\tilde{K}_a^i=e^b_i\tilde{K}_{ab}$.  Combining this with
(\ref{partial-a3}), (\ref{aexpansion}), (\ref{cbk1}), we find
$e^{b}_{j}k_{b}^{l}+e^{bl}k_{bj} = \kappa \gamma
\theta \delta^{l}_{j}J^{0}/2(1+\gamma^{2})$.  On the other hand, from
the Gauss constraint it follows that
$k_{b}^{j}e^{bl}-k_{b}^{l}e^{bj} = \kappa \gamma
\beta\epsilon_{i}^{\ jl}J^{i}/2(1+\gamma^{2})$.  Thus,
\begin{eqnarray}
\label{kam}
k^{m}_{a} = \frac{\kappa
  \gamma}{4(1+\gamma^{2})}\left(\beta \epsilon_{ij}^{\ \ m}\
e_{a}^{i}J^{j}+\theta e_{a}^{m}J^{0}\right)
\end{eqnarray}
is the contribution which provides the antisymmetric part of $K_{ab}$,
but also adds to the symmetric term.  The expression for $k^{m}_{a}$
can independently (but not fully canonically) be verified by computing
it from (\ref{CIJK}) as $k^{m}_{a}=
-C_{a}^{m0}=-q^{\nu}_{a}n_{J}C_{\nu}^{JM}$.

With (\ref{cbk1}) and
(\ref{kam}), the Ashtekar--Barbero connection as split into its
torsion and torsion-free parts is
\begin{eqnarray}
\label{abconnection1}
A_{a}^{i}=
\widetilde{\Gamma}^{i}_{a}+\gamma \widetilde{K}^{i}_{a}+\frac{\kappa
  \gamma}{4}\ \epsilon^{i}_{\ kl}e^{k}_{a}J^{l}-
\frac{\kappa \gamma}{4\alpha}e_{a}^{i}J^{0}\,, 
\end{eqnarray}
where the first term is completely torsion-free and only the $J$-terms
represent the torsion contribution.

\subsection{Parity transformation}

We first define the parity trasnformation for both canonical
gravitational variables and fermionic matter fields such that it
respects the background independence of a theory of gravity
non-minimally coupled with fermions. Parity conservation can then be
determined by testing either whether the effective action
(\ref{interactingaction1}) in the Lagrangian formulation, or
constraints as well as the symplectic structure of the Hamiltonian
formulation are left invariant.  As we will see, the torsion
contributions to the connection play an important role in this, and we
will be led to split all the constraints into their torsion-free and
torsion parts to verify the parity behavior.

In a background-independent setting, we cannot refer to spatial
coordinates changing their sign under parity reversal. Instead, as
usually in formulations on curved manifolds we use the fact that
triads change their orientations under parity reversal as one of the
primary contributions to the parity transformation: $e^a_i\rightarrow
- e^a_i$ \footnote{More generally, the triads can be allowed to
transform as $e^a_j\rightarrow \Lambda^i_j e^a_i$, where $\Lambda^i_j$
is an orthogonal transformation matrix with determinant $-1$. Also,
the gamma matrices transform like $\gamma^{0}\rightarrow
\gamma^{0}$and $\gamma^{i}\rightarrow - \ \Lambda^i_j\gamma^{j}$. It
is easy to check that the torsion-free spin connection and the
extrinsic curvature transform as $\widetilde{\Gamma}^i_a \rightarrow -
\Lambda^{i}_{j}\widetilde{\Gamma}^j_a$ and $\widetilde{K}^i_a
\rightarrow \Lambda^{i}_{j}\widetilde{K}^j_a$. Finally, the
transformation of $A_a^i$ can be obtained from these two
one-forms. Our arguments about parity invariance remain unchanged if
this more general transformation is used.}.  For Dirac spinors, we use
the conventional field theory definition $\Psi \rightarrow
\gamma^{0}\Psi$.  These basic definitions imply
\begin{eqnarray}
\label{ptransformations}
J^{0} &=& \overline{\Psi} \gamma^{0}\gamma^{5}\Psi \rightarrow - J^{0} \
\  , 
\ J^{i} = \overline{\Psi} \gamma^{i}\gamma^{5}\Psi \rightarrow J^{i}\\
\Gamma^i_a &=& \widetilde{\Gamma}^i_a + C^i_a \rightarrow 
\widetilde{\Gamma}^i_a - \frac{\gamma\kappa}{4(1+\gamma^{2})}
\left(\theta 
\ \epsilon_{ij}^{\ \  k}e_{b}^{i}J^{j}+\beta e_{b}^{k} J^{0}\right)
\nonumber\ \\
K^i_a &=& \widetilde{K}^i_a + k^i_a \rightarrow - \widetilde{K}^i_a - 
\frac{\gamma\kappa}{4(1+\gamma^{2})}
\left(\beta 
\ \epsilon_{ij}^{\ \  k}e_{b}^{i}J^{j}-\theta e_{b}^{k} J^{0}\right) \nonumber
\end{eqnarray}
where we have used
$\widetilde{K}^i_a=\widetilde{K}_{ab}e^{bi}\rightarrow -
\widetilde{K}^i_a$. The $\tilde{\Gamma}_a^i$-behavior follows from
formulas in the Appendix, which keep track of factors of ${\rm
sgn}\det (e_a^i)$. 

It is interesting to note that both $\Gamma^i_a$ and $K^i_a$ transform
as their torsion-free counterparts $\widetilde{\Gamma}^i_a$ and
$\widetilde{K}^i_a$ only for $\alpha = \gamma$ (i.e.\ $\theta=0$), a
result expected from \cite{FermionAshtekar}. Also note that the
Ashtekar-Barbero connection $A_a^i$ does not have a simple
transformation property because $\Gamma^i_a$ and $K^i_a$ transform
differently. However, as seen in (\ref{abconnection1}) the torsion
contributions simplify when combined to $C_a^i+\gamma
k_a^i$. Regrouping the remaining terms by new combinations with
$\widetilde{\Gamma}_a^i$ and $\widetilde{K}_a^i$ provides a
transformation law
\begin{widetext}
\begin{eqnarray}
\label{atransformation}
A^i_a = \left(\widetilde{\Gamma}^i_a - \frac{\gamma \kappa}{4\alpha}\ 
e_a^{i}J^{0}\right)+ \gamma\left(\widetilde{K}^i_a +\frac{\kappa}{4}\ 
\epsilon^{i}_{\ jk} e_a^{j}J^{k}\right)\rightarrow 
\left(\widetilde{\Gamma}^i_a - \frac{\gamma \kappa}{4\alpha}\ 
e_a^{i}J^{0}\right)- \gamma \left(\widetilde{K}^i_a +\frac{\kappa}{4}\ 
\epsilon^{i}_{\ jk} e_a^{j}J^{k}\right)
\end{eqnarray}
\end{widetext}
just like the combination of torsion-free $\widetilde{\Gamma}^i_a$ and
$\widetilde{K}^i_a$.

With these rules, the Liouville term in the action transforms as
\begin{widetext}
\begin{eqnarray}
\label{symptransformation}
\int_{\Sigma}\md^{3}x P^{a}_{i}{\cal L}_{t}A^i_a &=&
\int_{\Sigma}\md^{3}x P^{a}_{i}{\cal
L}_{t}\left(\left(\widetilde{\Gamma}^i_a - \frac{\gamma
\kappa}{4\alpha}\ e_a^{i}J^{0}\right)+ \gamma\left(\widetilde{K}^i_a
+\frac{\kappa}{4}\ \epsilon^{i}_{\ jk}
e_a^{j}J^{k}\right)\right)\nonumber\ \\ &\rightarrow&
-\int_{\Sigma}\md^{3}x P^{a}_{i}{\cal
L}_{t}\left(\widetilde{\Gamma}^i_a - \frac{\gamma \kappa}{4\alpha}\
e_a^{i}J^{0}\right)+ \gamma\int_{\Sigma}\md^{3}x P^{a}_{i}{\cal
L}_{t}\left(\widetilde{K}^i_a +\frac{\kappa}{4}\ \epsilon^{i}_{\ jk}
e_a^{j}J^{k}\right)=
\int_{\Sigma}\md^{3}x
P^{a}_{i}{\cal L}_{t}A^i_a,
\end{eqnarray}
\end{widetext}
where we have used the fact that
$\left\{P^{a}_{i},\widetilde{\Gamma}^i_a - \frac{\gamma
\kappa}{4\alpha}\ e_a^{i}J^{0}\right\}_{PB}=0$ such that the
$\Gamma$-term does not contribute to the symplectic
structure. Therefore, the symplectic structure is invariant under the
parity transformation.

Given that $A_a^i$ consists of two terms transforming differently, it
is useful for a parity analysis to rewrite all terms of the
constraints by explicitly splitting off their torsion
contributions. The torsion-free parts will then just have the vacuum
parity behavior, which is parity invariant, while the torsion terms
directly demonstrate the parity behavior in the presence of fermions
through the currents.  The Gauss constraint can easily be split in
this way and formulated in torsion-free variables. The split Gauss
constraint, $\widetilde{G}_{i}=\gamma \epsilon_{ij}^{\ \
k}\widetilde{K}_a^{j} P^{a}_{k} = 0$, is independent of the fermion
current and thus parity invariant. Splitting the diffeomorphism
constraint into torsion and torsion-free components is more involved,
and after a longer computation we obtain
\begin{widetext}
\begin{eqnarray}
\label{diff1}
C_{a}&=&
P^{b}_{j}(\widetilde{F}_{ab}^{j}+
2\partial_{[a}\overline{A}^{j}_{b]}+\epsilon^{j}_{\ 
  lm}\overline{A}^{l}_{a}\overline{A}^{m}_{b}+\epsilon^{j}_{\
  lm}\overline{A}^{l}_{a}\widetilde{A}^{m}_{b}+\epsilon^{j}_{\
  lm}\widetilde{A}^{l}_{a}\overline{A}^{m}_{b})-
\frac{1+\gamma^{2}}{\gamma}K_{a}^{i}G_{i} \nonumber\ \\ && 
-\frac{1}{2}i\sqrt{q}(\psi^{\dagger}\widetilde{D}_{a}{\psi}+
\eta^{\dagger}\widetilde{D}_{a}{\eta}- c.c.)-
\frac{1}{2}C_{a}^{i}\sqrt{q}J_{i}-\frac{\gamma}{2}{K}_{a}^{i}\sqrt{q}J_{i}\\
&=& 2\gamma P^{b}_{j}\widetilde{D}_{[a}\widetilde{K}_{b]}^{j}+{\rm sgn}\det (e_a^i)\frac{\gamma
  \kappa}{4}\epsilon_{ca}^{\ \ \
  b}{P}_{l}^{c}\widetilde{D}_{b}(\sqrt{q}J^{l})
  -\frac{1}{2}i\sqrt{q}(\psi^{\dagger}\widetilde{
  D}_{a}{\psi}+\eta^{\dagger}\widetilde{ D}_{a}{\eta}- c.c.)
  \nonumber\ \\
&&  +{\rm sgn}\det (e_a^i)P^{f}_{l}\left(\epsilon_{cf}^{\ \ b}
\Gamma^{c}_{ba}-\epsilon_{ca}^{\ \ b}\Gamma^{c}_{bf}\right)
\sqrt{q}J^{l}+\left(\frac{\gamma \kappa}{4}\epsilon^{jkl}J_{k}e_{al}-
\frac{\gamma \kappa}{4\alpha}e_{a}^{j}J^{0}-\frac{1+\gamma^{2}-
\gamma^{3}}{\gamma}K^{j}_{a}\right)\widetilde{G}_{j}\,,  \nonumber
\end{eqnarray}
\end{widetext}
where $\Gamma^{c}_{ab}$ is the torsion-free Christoffel connection
which can be expressed in terms of triads and co-triads as in
(\ref{spin1}) and we have used (\ref{abconnection1}) and
$\gamma[\widetilde{K}_{b},P^{b}]_{i}=\widetilde{G}_{i}$ to arrive at
the final expression. Again, the splitting makes it obvious that the
diffeomorphism constraint is invariant under parity
transformations. Notice the importance of ${\rm
  sgn}\det(e^i_a)$-factors which we carried through the calculation
--- see also the Appendix for some formulas.

Finally, the Hamiltonian constraint turns out to be
\begin{widetext}
\begin{eqnarray}
\label{hamiltonian1}
C\ &=& \ \frac{\kappa
  \gamma^{2}}{2\sqrt{q}}{P}^{a}_{i}P^{b}_{j}\left(\epsilon^{ij}_{\ \
    k}(\widetilde{F}_{ab}^{k}+2\partial_{[a}\overline{A}^{k}_{b]}+
\epsilon^{k}_{\ lm}\overline{A}^{l}_{a}\overline{A}^{m}_{b}+
\epsilon^{k}_{\ lm}\overline{A}^{l}_{a}\widetilde{A}^{m}_{b}+
\epsilon^{k}_{\ lm}\widetilde{A}^{l}_{a}\overline{A}^{m}_{b})-
2(\gamma^{2}+1)K_{[a}^{i}K_{b]}^{ j}\right)\nonumber\ \\  
&&+\frac{1}{2}i\gamma
  \kappa{P}^{a}_{i}(\psi^{\dagger}\sigma^{i}\widetilde{D}_{a}\psi-
  \eta^{\dagger}\sigma^{i}\widetilde{D}_{a}\eta-c.c.)
  +\frac{\gamma^{2} \kappa
  {P}_{i}^{a}}{2\sqrt{q}}\widetilde{D}_{a}(\sqrt{q}J^{i})
  +\frac{\gamma^{2}\kappa}{2}{P}_{j}^{b}K_{b}^{j}J^{0} 
 \nonumber\ \\ &&+ \frac{\gamma \kappa}{2}[K_{a},P^{a}]_{j}J^{j}
  -\frac{3\kappa}{8\sqrt{q}}\frac{\gamma^{2}}{1+\gamma^{2}}q(J_{0})^2+
  (1+\gamma^{2})\kappa\widetilde{D}_{a}
\left(\frac{{P}_{i}^{a}G^{i}}{\sqrt{q}}\right)\\
\label{hamiltonianconstraint}
 &=& \ \frac{\kappa
  \gamma^{2}}{2\sqrt{q}}{P}^{a}_{i}P^{b}_{j}\left(\epsilon^{ij}_{\ \
    k}\widetilde{R}_{ab}^{k}-2\widetilde{K}_{[a}^{i}
  \widetilde{K}_{b]}^{ j}\right)+\frac{1}{2}i\gamma
  \kappa{P}^{a}_{i}(\psi^{\dagger}\sigma^{i}{\partial}_{a}\psi-
  \eta^{\dagger}\sigma^{i}{\partial}_{a}\eta-c.c.) + \frac{\gamma
  \kappa 
\theta}{2}P^{b}_{j}\widetilde{\Gamma}^{j}_{b}J^{0}+\frac{\gamma^{3}
\kappa^{2}}{4\alpha\sqrt{q}}\epsilon^{ij}_{\ \
  k}P^{a}_{i}e^{k}_{b}J^{0}
\partial_{a}P^{b}_{j}\nonumber\ \\ &&   
 +\frac{3\kappa}{16}\frac{\gamma^{2}}{1+\gamma^{2}}
\left(\frac{1}{\alpha^{2}}-\frac{2}{\alpha\gamma}-1\right)\sqrt{q}(J_{0})^2
 -\frac{3\kappa}{16}\frac{\gamma^{2}}{1+\gamma^{2}}
\left(\frac{1}{\alpha^{2}}-\frac{2}{\alpha\gamma}-1\right)\sqrt{q}J_{l}J^{l}
+\kappa\widetilde{D}_{a}\left(\frac{{P}_{i}^{a}
  \widetilde{G}^{i}}{\sqrt{q}}\right)+ 
  \frac{\kappa}{2}\left(1+\frac{\gamma^{2}}{2}\right)\widetilde{G}_{i}J^{i}
\nonumber
\end{eqnarray}
\end{widetext}
where $\widetilde{R}_{ab}^k$ is the curvature of
$\widetilde{\Gamma}_a^i$.  Also this expression is parity invariant
for all $\alpha$.

Independently of the parity behavior, the last expression allows
us to provide a cross-check of our constraints by comparing with the
effective action (\ref{interactingaction1}). From the Hamiltonian
constraint (\ref{hamiltonian1}) we read off the
interaction term
\begin{widetext}
\begin{eqnarray}
\label{interactinghamiltonian}
H_{\rm int}\ = \
\frac{3\kappa}{16}\frac{\gamma^{2}}{1+\gamma^{2}}
\left(\frac{1}{\alpha^{2}}-\frac{2}{\alpha\gamma}-1\right)\sqrt{q}(J_{0})^2
-\frac{3\kappa}{16}\frac{\gamma^{2}}{1+\gamma^{2}}
\left(\frac{1}{\alpha^{2}}-\frac{2}{\alpha\gamma}-1\right)
\sqrt{q}J_{l}J^{l}\,, 
\end{eqnarray}
\end{widetext}
quadratic in currents. This agrees with the Lagrangian formulation
(\ref{interactingaction1}). In addition to checking the parity
behavior, splitting the constraints into torsion-free/torsion parts
thus provides a non-trivial cross-check by comparing our constraints
with the interaction Hamiltonian of the effective action \footnote{This
demonstrates that terms presented here, and which differ from those in
\cite{FermionAshtekar} (for $\alpha=\gamma$), must be contained in the
constraints.  Ignoring the interaction term in
(\ref{hamiltonianconstraint}), on the other hand, provides the
Hamiltonian constraint of a second-order formalism which can be
compared directly with the Appendix of \cite{FermionHiggs} (for
$\gamma=1$). Notice that the derivation sketched in
\cite{FermionHiggs} does not work purely in real variables and assumes
properties of the projection from complex variables. As the comparison
with our results shows, the calculations of \cite{FermionHiggs} leave
some extra terms in the constraint which are absent in a complete
derivation based only on real variables.}.

\section{Quantization}

In Sec.~\ref{sec:canonical Formulation}, we have summarized all the
necessary generalizations in the canonical formulation which are
induced by the coupling between gravity and fermions through the
Ashtekar--Barbero connection, and explicitly verified parity
invariance in Sec.~\ref{s:Parity} which is not manifest in canonical
variables $(A_a^i,P^b_j)$.  In this section, we explore the effects on
the canonical quantization as used in loop quantum
gravity. Quantizations of fermions in canonical gravity have already
been developed in \cite{QSDV,FermionHiggs} (see also
\cite{LoopFermion1,LoopFermion2,LoopFermion3} for earlier work).
However, these developments were not based on a systematic derivation
of the canonical formulation from a Holst-type action and several
features related to the torsion-dependence of the Ashtekar--Barbero
connection were overlooked or remained implicit.  Corresponding
adaptations which become necessary in a consistent quantization could
thus appear to draw suspicions about the validity of the basic
strategy of a loop quantization as used in
\cite{QSDV,FermionHiggs}. (Some concerns have, for instance, been
voiced in \cite{NPZRev}.) In addition to that, we here raise the
question of parity invariance of the quantum theory which may be a
concern given that the classical verification of parity required us to
partially solve equations of motion to see the correct transformation
behavior.

Before starting the quantization, the first question concerns the
choice of basic variables. We have two sets, given by the canonical
variables $(A_a^i,P^b_j)$ in the presence of torsion as well as the
torsion-free components $(\tilde{A}_a^i,P^b_j)$ with explicit
expressions for torsion in terms of the fermion current in
(\ref{abconnection1}). However, as we have seen, equations of motion
are required to find the torsion contribution to extrinsic curvature
in explicit form. The use of classical equations of motion is not
suitable for a quantization, and there is thus no choice but to use
the canonical variables with implicit torsion terms.

\subsection{Half-densitized fermions}

In addition to torsion terms, there will be a further contribution to
the connection once we formulate the fermions in terms of
half-densities as required for consistency \cite{FermionHiggs}.  For
fermions, we have the canonical pair $(\psi,\pi)$ with
$\pi=-i\sqrt{q}\psi^\dagger$. These canonical variables
cannot be promoted to operators on a Hilbert space with a suitable
inner product in a way incorporating the reality condition
$\pi^{\dagger}=i\sqrt{q}\psi$ by satisfying
$\hat{\pi}^{\dagger}=i\widehat{\sqrt{q}}\hat{\psi}$: First,
if $f(A)$ is a non-trivial real valued function of the connection $A$,
then the inconsistent relation
\begin{equation}
\label{inconsistency}
0=0^{\dagger}=([\hat{\pi},f(A)])^{\dagger}=
i[\widehat{\sqrt{q}},f(A)]\hat{\psi}\neq 0
\end{equation}
ensues.  Here the first commutator is expected to vanish since the
corresponding classical Poisson bracket vanishes. On the contrary, the
classical Poisson bracket corresponding to the second commutator is
non-zero; hence the inconsistency arises. A second problem can be seen
to arise from the symplectic structure obtained from the fermion Liouville
form
\begin{widetext}
\begin{eqnarray}
\label{inconsisteny1}
\Theta =  -i\int_{\Sigma_{t}} \md^{3}x 
\sqrt{q}\left(\theta_{L}\psi^{\dagger}\dot{\psi}-\theta_{R}
\dot{\psi^{\dagger}}\psi\right)
=  \int_{\Sigma_{t}} \md^{3}x \left(
\pi
  \dot{\psi}-\frac{i}{2}\theta_{R}\gamma\kappa {\psi^{\dagger}}\psi
  e^{i}_{c}\dot{P}_{i}^{c}\right)-\int_{\Sigma_{t}} \md^{3}x 
\theta_R{\cal
    L}_{t}(\pi \psi)\,.
\end{eqnarray} 
\end{widetext}
Here, it follows from the second term of the first integral that the
connection $A_{a}^{i}$ acquires an imaginary correction term
$\frac{1}{2}i\theta_R{\psi^{\dagger}}\psi e^{i}_{a}$, which endows the
theory with a complex connection. This, in turn, would require the use
of a complexification of SU(2) in holonomies, for which, due to the
non-compactness, none of the loop quantization techniques relying on
the existence of a normalized Haar measure would be available (see
e.g.\ \cite{ALMMT}).

Both problems were solved by Thiemann who observed in
\cite{FermionHiggs} that, in order to obtain a well-defined canonical
loop quantization with a real Ashtekar--Barbero connection also in the
presence of fermions, one should cast fermion fields into
Grassmann-valued half-densities. Thus $\xi:=\sqrt[4]{q}\psi$ instead
of $\psi$ (and $\chi:=\sqrt[4]{q}\eta$ instead of $\eta$) is
considered to be the classical canonical variable, and $\pi_{\xi}=
-i\xi^{\dagger}$ is the conjugate momentum for $\xi$.  The
inconsistencies in (\ref{inconsistency}) are naturally removed as the
new canonical variables imply the reality condition
$\pi_{\xi}^{\dagger}=i\xi$ without any appearance of $\sqrt{q}$.

In half-densities, the symplectic structure becomes
\begin{widetext}
\begin{eqnarray}
\label{symplectomorphism}
\Theta \ &=& \ -i	\int_{\Sigma_{t}} \md^{3}x \
\sqrt{q}\left(\theta_{L}(\psi^{\dagger}\dot{\psi}-\dot{\eta^{\dagger}}\eta)-
\theta_{R}(\dot{\psi^{\dagger}}{\psi}-\eta^{\dagger}\dot{\eta})\right) 
= \int_{\Sigma_{t}} \md^{3}x
\left(\pi_{\xi}\dot{\xi}+\pi_{\chi}\dot{\chi}
\right)+
\int_{\Sigma_{t}} \md^{3}x \ \frac{\gamma \kappa}{4\alpha}
P^{c}_{i}{\cal L}_{t}(e_{c}^{i}J^{0})\,, 
\end{eqnarray}
\end{widetext}
where we have ignored total time derivatives which would drop out of
the action for appropriate boundary conditions.  The classical
anti-Poisson brackets for Grassmann-valued fields are
$\left\{\xi_{A}(x),\pi_{\xi B}(y)\right\}_{+}=\delta_{AB}
\delta(x,y)$. Moreover, as the extra term shows, $\sqrt[4]{q}$ can
be absorbed in spinors without changing the symplectic structure of
the gravitational variables only when $\alpha\to\infty$, i.e.\ for
minimal coupling.  Combining the last term in
(\ref{symplectomorphism}) with the gravitational Liouville term
$\int\md^3x P^c_i{\cal L}_tA_c^i$, a real-valued correction term
$\frac{\gamma \kappa}{4\alpha}e_{a}^{i}J^{0}$ must be added to the
Ashtekar-Barbero connection $A_{a}^{i}$. This is a new feature that is
present in the non-minimally coupled theory if the fermion fields are
expressed in terms of half-densities. Therefore, the new canonical
connection can be written as
\begin{eqnarray}
\label{correctedconnection}
{\cal A}_{a}^{i}\ := \ A_{a}^{i} + \frac{\gamma \kappa}{4\alpha}e_{a}^{i}
J^{0} \ =\widetilde{\Gamma}_{a}^{i} + {\cal C}_{a}^{i}+\gamma K_{a}^{i}\, ,
\end{eqnarray}
where 
\begin{eqnarray}
\label{correctedc}
{\cal C}_{a}^{i}\ := \frac{\theta\gamma^{2}
  \kappa}{4(1+\gamma^{2})}
\left(\frac{1}{\gamma} \ \epsilon^{j}_{\
    kl}e_{a}^{k}J^{l}-  e_{a}^{j} J^{0}\right).
\end{eqnarray}
Absorbing the correction term into the torsion contribution to the
spatial spin connection allows one to keep $K_{a}^{i}$ unchanged in
the course of expressing all the constraints in terms of the corrected
connection. Note that the corrected torsion contribution, ${\cal
C}_{a}^{i}$, to the spin connection vanishes for $\alpha=\gamma$. (If
one would use the fully split connection (\ref{abconnection1}) based
on partial solutions of the equations of motion, the new contribution
in the presence of half-densities would cancel the $J^0$-dependence of
${\cal A}_a^i$ completely.)

In terms of the corrected connection and half-densities, the total
Dirac Hamiltonian constraint (modulo the Gauss constraint) in
(\ref{totalhamiltonian}) takes the smeared form
\begin{widetext}
\begin{eqnarray}
\label{hdhamiltonianconstraint}
H_{\rm total}&=& \int_{\Sigma_{t}} \md^{3}x \ N\left(
\frac{\gamma^{2}\kappa}{2\sqrt{q}}{P}^{a}_{i}P^{b}_{j}\left(\epsilon^{ij}_{\
  \ k}{\cal F}_{ab}^{k}-2(\gamma^{2}+1)K_{[a}^{i}K_{b]}^{ j}\right)
-\frac{\gamma\kappa\beta
   P^{a}_{i}}{\sqrt{q}}{\cal
   D}_{a}\left(\pi_{\xi}\tau^{i}\xi+\pi_{\chi}
\tau^{i}\chi\right)\right.\nonumber\\
&&-i\frac{2\gamma \kappa P^{a}_{i}}{\sqrt{q}} 
\left(\theta_{L}\pi_{\xi}\tau^{i}{\cal D}_{a}\xi 
 -\theta_{R}\pi_{\chi}\tau^{i}{\cal D}_{a}\chi -
 c.c.\right)
+\frac{\gamma \kappa \beta}{2\sqrt{q}(1+\gamma^{2})}\left(3-
\frac{\gamma}{\alpha}+
2\gamma^{2}\right)(\pi_{\xi}\tau_{l}\xi+\pi_{\chi}
\tau_{l}\chi)(\pi_{\xi}\tau^{l}\xi+\pi_{\chi}
\tau^{l}\chi)\nonumber\ \\
&& +i\frac{\gamma^{3}\kappa^{2}}{4\alpha{q}}\epsilon^{ij}_{\ \ k}
P^{a}_{i}e_{b}^{k}(\pi_{\xi}\xi-\pi_{\chi}\chi)
{\cal D}_{a}P^{b}_{j}+\left.\frac{3\gamma\kappa\theta}{8\alpha \sqrt{q}}
(\pi_{\xi}\xi-\pi_{\chi}\chi)(
\pi_{\xi}\xi-\pi_{\chi}\chi)\right)
\end{eqnarray}
\end{widetext}
where ${\cal F}_{ab}^{k}$ is the curvature and ${\cal D}$, now and in
the rest of the paper, is the covariant derivative related to the
corrected connection ${\cal A}$.

\subsection{Quantum representation}

The ordinary kinematical constructions of loop quantum gravity do not
refer to torsion or torsion-freedom and thus go through unchanged. We
thus present only the bare concepts relevant for the construction of
constraint operators.

\subsubsection{Fermion fields}

The space of all Grassmann-valued half-densitized 2-component spinors
$\xi(x)$ and $\chi(x)$ constitutes the classical configuration space
$\cal F$ for fermion fields. The loop quantization \cite{FermionHiggs}
then promotes smeared objects
\[
\Xi_{A}(x) := \ \int_{\Sigma_{t}} \md^{3}y
\sqrt{\delta(x,y)}\xi_{A} \ := \lim_{\epsilon \rightarrow
  0}\int_{\Sigma_{t}} {\md}^{3}y
\sqrt{\frac{\chi_{\epsilon}(x,y)}{{\epsilon^{3}}}}\xi_{A} 
\]
to operators, where $\chi_{\epsilon}(x,y)$ is the characteristic
function of a box of Lebesgue measure $\epsilon^{3}$ centered at
$x$. Note that $\Xi_A$ are scalar Grassmann valued functions since the
$\delta$ distribution is a density of weight one. It is also easy to
see that $\Xi$ and their adjoint satisfy anti-Poisson brackets similar
to those presented above for $\xi$.  Upon quantization, the
anti-Poisson bracket is replaced by the anti-commutator
$[\hat{\Xi}_A(x),\hat{\pi}_B(y)]_{+} = i\hbar \delta_{AB}\delta_{x,y}$
with $\delta_{x,y}$ being the Kronecker symbol (rather than a
$\delta$-distribution thanks to the smearing involved in $\Xi_A$).

This algebra can be represented on a non-separable Hilbert space
${\cal H}_F=L^{2}(\overline{S},\md\mu_{F})=\bigotimes_{v\in
\Sigma}L^{2}({S_{v}},\md \mu_{v})$ where each copy ${\cal H}_v$ for
any point $v$ in space is an ordinary Grassmann-valued Hilbert space
of multi-linear functions of $\Xi_A(v)$ and $\bar{\Xi}_A(v)$ of
two-component spinors in their Grassmann space $S_v$, with integration
measure $\md \mu_v= \md\overline{\Xi}_v\md\Xi_v
e^{\overline{\Xi}_v\Xi_v}$. The full space of the fields can then be
written as $\overline{S}:=\bigotimes_{v\in \Sigma}S_{v}$ with measure
$\md\mu_{F}(\overline{\Xi},\Xi)=\prod_{v\in \Sigma}\md \mu_{v}$. On
this space, $\hat{\Xi}_A$ acts as a multiplication operator, and its
momentum $\hat{\pi}_B=-i\hbar \partial/\partial\Xi_B$ by a derivative.
In addition, we have a second copy of these point-wise Hilbert spaces
for $\chi$ smeared to $X$.

A dense subset of functions in this Hilbert space is formed by
cylindrical functions which are superpositions only of products of
finitely many vertex-wise Grassmann-factors. These functions can be
seen to arise if one starts with a cyclic state independent of $\Xi$
and $X$ and uses the $\hat{\Xi}_v$ and $\hat{X}_v$ as ``creation''
operators. Since all the constraints depend on the fermion only via
currents, which are polynomials in $\Xi_A$ and $X_A$, they can easily
be represented on this subspace of cylindrical functions.

\subsubsection{Gravitational variables}

Classical configuration variables for gravity are SU(2)-connections on
a principal fiber bundle over the spatial manifold $\Sigma$,
represented by smooth su(2)-valued local 1-forms ${\cal A}^{i}_{a}$
from (\ref{correctedconnection}); the space ${\cal A}$ of all such
1-forms is the classical configuration space. The phase space is
coordinatized by the pair $({\cal A}^{i}_{a},P^{a}_{i})$, where
$P^{i}_{a}$ is the conjugate momentum, an su(2)-valued vector density
on $\Sigma$ proportional to the densitized triad. Then the only
non-vanishing Poisson bracket is
\begin{equation}
\label{poissonb}
\left\{{\cal A}^{i}_{a}(x),P^{b}_{j}(y)\right\} =  
\delta^{i}_{j} \delta^{b}_{a} \delta (x,y)\,.
\end{equation}
No well-defined quantum analogs for these canonical variables exist in
a direct form without smearing.  The elementary classical variables
that have well-defined quantum analogs are rather given by (complex
valued) matrix elements of holonomies $h_{e}({\cal A})={\cal
P}\exp(\int_e{\cal A}_a^i\tau_i\dot{e}^a\md t)\in {\rm SU}(2)$ along
paths $e$ in $\Sigma$ and fluxes $F_S^{(f)}(P):=
\int_{S}f_{i}n_aP^a_i\md^2y$, where $f$ are su(2)-valued functions
across 2-surfaces $S$ in $\Sigma$ and $n_a$ is the
(metric-independent) co-normal to the surface.

This provides the appropriate smearing for gravitational
variables. The resulting holonomy-flux algebra is represented on a
Hilbert space ${\cal H}=L^2(\overline{{\cal A}},\md\mu_{\rm AL})$
constructed as follows \cite{ALMMT}: We first introduce cylindrical
functions whose space will eventually be completed to a Hilbert space.
Cylindrical functions are functions on $\cal A$ which depend on ${\cal
A}_a^i$ only through holonomies $h_{e}(A)$ along edges $e$ of a graph
$\alpha$ (a finite set of edges) in $\Sigma$. If a graph $\alpha$ has
$n$ edges, then, given a $C^{\infty}$ complex-valued function $\psi$
on ${\rm SU}(2)^{n}$, a cylindrical function $\Psi_{\alpha}$ on $\cal
A$ with respect to the graph $\alpha$ can be written as
\begin{equation}
\label{cylfunc}
\Psi_{\alpha}({\cal A}):= \psi(h_{e_{1}}({\cal
  A}),\ldots,h_{e_{n}}({\cal A}))\,.
\end{equation}
Let $\rm{Cyl}_{\alpha}$ denote the space of such functions with
respect to the graph $\alpha$, and $\rm{Cyl}=
\cup_{\alpha}\rm{Cyl}_{\alpha}$ the space of all cylindrical
functions. A natural inner product on $\rm{Cyl}_{\alpha}$ can be
introduced by defining the measure $\md\mu_{\alpha}$ by
\begin{widetext}
\begin{equation}
\label{innerproduct}
\left\langle \Psi_{\alpha},\Phi_{\alpha}\right\rangle \ = \ 
\int d\mu_{\alpha}\overline{\psi}_{\alpha}\phi_{\alpha}:=
\int_{{\rm SU}(2)^n} \md\mu_{\rm H}^n \overline{\psi_{\alpha}(h_{e_1},
\ldots,h_{e_n})}\phi_{\alpha}(h_{e_1},
\ldots,h_{e_n})
\end{equation}
\end{widetext}
with the Haar measure $\md\mu_{\rm H}$ on SU(2). The
Cauchy completion of $\rm{Cyl}_{\alpha}$ with respect to this inner
product gives rise to a Hilbert space ${\cal H}_{\alpha}:=
{L^{2}}({\overline{\cal A}}_{\alpha},{\md \mu_{\alpha}})$, where
$\overline{\cal A}_{\alpha}:={{\cal A}_{\alpha}} / {\cal
G}_{\alpha}^{0}$ is the space of smooth connections restricted to
the graph $\alpha$ modulo all local gauge transformations $g_{\alpha}
\in {\cal G}_{\alpha}^{0}$ which are the identity on the vertices.
 
The measure extends to the full space ${\rm Cyl}$ (where spaces ${\rm
  Cyl}_{\alpha}$ for different $\alpha$ are mutually orthogonal except
  for states which can be written cylindrically with respect to both
  graphs) and, by completion, defines the full Hilbert space ${\cal
  H}:= {L^{2}}({\overline{\cal A}},{\md \mu_{\rm AL}})$ where
  $\md\mu_{\rm AL}$ is the Ashtekar--Lewandowski measure constructed
  in this way and $\overline{\cal A}$ the space of generalized
  connections. The latter space represents the quantum configuration
  space as an enlargement from the classical configuration space $\cal
  A$ of connections by distributions. On ${\cal H}$, holonomies act as
  multiplication operators which change the graph when acting on a
  cylindrical state whose graph does not contain the edge used in the
  holonomy. Flux operators are represented by invariant vector field
  operators on SU(2)-copies corresponding to the edges intersected by
  the surface of the flux. As operators on function spaces over SU(2),
  invariant vector fields have discrete spectra, and so do flux
  operators. From fluxes, one can construct further operators of
  spatial geometry such as area and volume \cite{AreaVol,Area,Vol2}
  which also have discrete spectra.

All this remains unchanged in the presence of torsion. By
construction, the Ashtekar-Barbero connection inherits the total
torsion contribution and thus the effect of torsion on the system is
concealed in holonomies which are used in states and as basic
multiplication operators of loop quantum gravity.  Consequently, the
functions of connections that represent the quantum state of the
system and operators containing holonomies are endowed with all
contributions from torsion in quantum kinematics. A complete split of
torsion-free and torsion components is possible only once equations of
motion are partially used. This is not available at the kinematical
level, which thus has no choice but to refer to the unsplit torsion
connection.

\subsubsection{Combined Hilbert space of gravity and fermions}

For the combined system, we simply take the tensor product ${\cal
  H}\otimes {\cal H}_F$ as the Hilbert space, which acquires the
  tensor product of the basic representations. All cylindrical states
  can be written in the form $\psi(h_{e_1},\cdots,
  h_{e_m},\Xi_{v_1},\cdots,\Xi_{v_m},X_{w_1},\cdots,X_{w_l})$ with
  integer $n$, $m$ and $l$. Especially for the gravitational
  dependence it is useful to use special cylindrical states based on
  spin networks \cite{RS:Spinnet,SpinNet}: graphs together with a
  labeling $j_e$ of their edges by irreducible SU(2)-representations
  $\rho^{(j_e)}$, and of vertices with spinor representations
  $\sigma_v$ of SU(2) (obtained from tensor products of the
  fundamental representation given by the basic 2-spinors) as well as
  contractors $C_v$ in vertices to contract the matrix-represented
  holonomies of edges incoming and outgoing at $v$. Such states take
  the form
\begin{equation} \label{spinnet}
 \prod_{v,e}
C_v^{\nu^v_1,\ldots,\nu^v_{n_v},\nu^v}{}_{\mu^v_1,\ldots,\mu^v_{m_v},\mu^v}
\rho^{(j_e)}(h_e({\cal A}))^{\mu_e}_{\nu_e}
\sigma_v(\Xi_v,X_v)^{\mu^v}_{\nu^v}
\end{equation}
where for all vertex labels $\nu^v_i$ are to be contracted with
indices $\nu_e$ on represented matrices $\rho^{(j_e)}(h_e({\cal
A}))^{\mu_e}_{\nu_e}$ of all $n_v$ outgoing edges as well as the
spinor index $\nu^v$, and $\mu^v_i$ with indices $\mu_e$ of all $m_v$
incoming edges as well as the spinor index $\mu^v$.

\subsection{Constraints}

General relativity is a background independent theory and is fully
constrained in the canonical formulation. Thus the quantization of the
constraints is necessary to obtain physical states. Having identified
elementary operators and their quantum representation, this
kinematical structure is now used to construct a set of quantum
operators corresponding to constraints relevant for the system.
Subsequently, these quantum constraints have to be solved to obtain
physical states. The existence of torsion may change the form of each
of the quantum constraint operators and consequently influence their
solutions. Here, we will show that extra terms can be quantized
consistently.
 
\subsubsection{Kinematical constraints}

We first express the Gauss constraint in terms of half-densities and
the new canonical connection ${\cal A}_a^i$:
\begin{equation}
\label{gc}
G_{i} := \ {\cal D}_{b}P^{b}_{i}-\frac{1}{2}\sqrt{q} J_{i}\ = \
{\cal D}_{b}P^{b}_{i} +\pi_{\xi}\tau_{i}\xi+ \pi_{\chi}\tau_{i}\chi\,.
\end{equation}
Upon smearing the constraint with an su(2)-valued function
$\Lambda^{i}$ on $\Sigma$, it is easy to see that
$G[\Lambda^i]=\int_{\Sigma} \md^{3}x \ \Lambda^{i}G_{i}$ generates
internal SU(2) rotations on the phase space of general relativity:
\[
\left\{{\cal A}_{a}^{i},G[\Lambda]\right\} = -{\cal
D}_{a}\Lambda^{i} \quad {\rm{and}} \quad \left\{P^{a}_{i},
G[\Lambda]\right\} = \epsilon_{ij}^{\ \ k}\Lambda^{j} P^{a}_{k}
\]
together with a spinor transformation in the fundamental
representation of SU(2). Thus, the quantization of the Gauss
constraint is carried out in a similar fashion as it is done in the
torsion-free case, restricting gauge invariant states to be supported
on ${\cal A}/{\cal G}$.  For our configuration variables, we have the
transformations $h_e\mapsto g_{e(0)}h_e g_{e(1)}^{-1}$, $\Xi_v\mapsto
g_v\Xi_v$ and $X_v\mapsto g_vX_v$ under a gauge transformation
$g\colon v\mapsto g_v\in {\rm SU}(2)$. A spin network state, when
gauge transformed, acquires at each vertex $v$ factors of
$\rho^{(j_e)}(g_v^{-1})$ from all incoming edges, $\rho^{(j_e)}(g_v)$
from outgoing edges and $f_v(g_v)$ from spinor factors in the
state. For a gauge invariant state, these factors must cancel each
other when contracted with the $C_v$ in (\ref{spinnet}), which implies
that representation matrices (including the spinor) must be multiplied
by contraction with an intertwiner of all relevant representations to
the trivial one.  The resulting gauge invariant states satisfy the
quantum constraint equation $\hat{G}[\Lambda^i]\Psi_{\alpha}=0$ for
all $\Lambda^i$.

Similarly, one can use the action of the spatial diffeomorphism group
on the phase space by computing infinitesimal canonical
transformations generated by $D[N^a]= \ \int_{\Sigma} \md^{3}x \
N^{a}C_{a}$. In terms of half-densities and the corrected connection,
the constraint turns out to be
\begin{widetext}
\begin{eqnarray}
\label{smeardc}
D[N^a] &=& \ \int_{\Sigma} \md^{3}x \ N^{a} \left(2
P^{b}_{j}\partial_{[a}{\cal A}_{b]}^{j}- {\cal
A}_{a}^{i}\partial_{b}P^{b}_{i}+
\frac{1}{2}(\pi_{\xi}{\partial}_{a}{\xi}-(\partial_a\pi_{\xi})\xi
+\pi_{\chi}{\partial}_{a}{\chi}- (\partial_a\pi_{\chi}) \chi)\right)
\end{eqnarray}
\end{widetext}
up to contributions from the Gauss constraint. This constraint
generates transformations
\[
\left\{{\cal A}_{a}^{i},D[N^a]\right\}=
N^{b}{\cal F}_{bc}^{i} + {\cal D}_{a}(N^{c}{\cal A}_{c}^{i}) = {\cal
L}_{\vec{N}}{\cal A}_{a}^{i}
\]
and
\[
\left\{P^{a}_{i},D[N^a]\right\}=
N^{b}\partial_{b}P^{a}_{i} -
P^{b}_{i}\partial_{b}N^{a}+P^{a}_{i}\partial_{b}N^{b}
= {\cal
L}_{\vec{N}}P^{a}_{i}
\]
as well as the correct Lie derivative $\delta\xi= N^a\partial_a\xi+
\frac{1}{2}\xi\partial_aN^a$ of half-densitized fermions.  Hence, this
constraint can be quantized as in the torsion-free case via the finite
action of the diffeomorphism group.  A finite diffeomorphism $\varphi$
is represented on cylindrical states by
\begin{widetext}
\begin{equation}
 \hat{D}_{\varphi}\psi(h_{e_1},\ldots,h_{e_m},
\Xi_{v_1},\ldots,\Xi_{v_n},X_{w_1},\ldots,X_{w_l})=
\psi(h_{\varphi(e_1)},\ldots, h_{\varphi(e_m)},
\Xi_{\varphi(v_1)},\ldots,\Xi_{\varphi(v_n)},X_{\varphi(w_1)},
\ldots,X_{\varphi(w_l)})
\end{equation}
\end{widetext}
simply by moving the graph (which presents a unitary transformation
with respect to the Ashtekar--Lewandowski measure). Thus, invariant
states can be determined by constructing a new, diffeomorphism
invariant Hilbert space via group averaging.

\subsubsection{Hamiltonian constraint}

While the Gauss and diffeomorphism constraints generate the canonical
transformations that represent the well-known kinematical gauge
symmetries in the classical phase space independently of torsion, the
scalar constraint entails the essence of dynamics of the theory. Hence
the scalar quantum operator describes quantum dynamics of the physical
states which must be in accordance with the presence of
torsion. Unfortunately, a complete quantization of this scalar
constraint is yet to be satisfactorily realized. Therefore, we present
only the necessary adaptations to the existing quantization
attempts. In this approach, it is essential to re-express the
classical expression of the scalar constraint in terms of those phase
space functions which can be promoted to well-defined operators.

Our starting point is expression (\ref{hdhamiltonianconstraint}) of
the Hamiltonian constraint in half-densitized fermions.  The fermion
terms in the Dirac Hamiltonian coupled with gravity, can be quantized
using the strategy developed by Thiemann in \cite{QSDV}. Note that
this Dirac Hamiltonian is different from the one presented in
\cite{QSDV} (which took a second order viewpoint) in two aspects: the
covariant derivative $\cal D$ now contains the Ashtekar--Barbero
connection with torsion and the interaction term is new. Also the
gravitational term has torsion contributions which have to be taken
into account when applying the standard quantization strategy of
\cite{QSDI}.

As usually, the expression involving extrinsic curvature $K_{a}^{i}$
would vanish for $\gamma=1$ in Euclidean signature which in turn
implies that the first term in the gravitational constraint
reduces to the scalar constraint $H^{E}[N]$ of Euclidean general
relativity. Then let us write the scalar constraint for gravity alone
as
\begin{equation}
\label{scforg}
H[N]=\sqrt{\gamma}H^{E}[N]-2(1+\gamma^{2}){\cal T}[N],
\end{equation}
where 
\begin{equation}
\label{T}
{\cal T}[N] := \sqrt{\frac{\gamma}{4\kappa}}\int_{\Sigma_{t}} \md^{3}x
\ N \frac{{P}^{a}_{i}P^{b}_{j}}{\sqrt{|\det P|}}
K_{[a}^{i}K_{b]}^{ j}\,. 
\end{equation}

In order to quantize the scalar constraint for gravity, it is first
necessary to express it in terms of classical phase space functions
which have well-defined quantum analogs. In this regard, the following
classical objects and relationships are crucial as building blocks:
The total volume $V= (\gamma\kappa)^{3/2}\int_{\Sigma} \md^{3}x
\sqrt{|\det P|}$ of $\Sigma$, the co-triad
\begin{equation}
\label{identity1}
e_{a}^{i}(x) := \frac{\sqrt{\kappa \gamma}}{2}
{\rm sgn}\det(e_d^l)\epsilon_{abc}\epsilon^{ijk}
\frac{{P}^{b}_{j}P^{c}_{k}}{\sqrt{{\rm{det}}\ P}}= \frac{2}{\gamma
  \kappa}\left\{{\cal A}_{a}^{i}(x),V\right\}\,,
\end{equation}
the integrated trace of extrinsic curvature
\begin{equation}
\label{K}
K \ := \ \gamma \kappa \int_{\Sigma} \md^{3}x \ K_{a}^{i}P^{a}_{i}\,.
\end{equation}
as well as expansions
\begin{eqnarray}
 h_e({\cal A}) &= & 1+\delta s^a\tau_i {\cal A}^i_a+O(\delta^2) \\
 h_{\alpha_{IJ}}({\cal A}) &=& 1+ \delta^2 s_I^as_J^b {\cal
 F}_{ab}^k\tau_k+O(\delta^3)
\end{eqnarray}
of holonomies along small open edges $e$ in direction $s^a$ of
coordinate length $\delta$ or small square loops $\alpha_{IJ}$ of
coordinate area $\delta^2$ with sides in the directions $s_I^a$.

The first step in a regularization of a spatial integral is to
introduce a triangulation of $\Sigma$ as the union of tetrahedra with
edges of coordinate length $\delta$ and edges at a given vertex
pointing in directions $s_I^a$, $I=1,2,3$. To use this for a
construction of operators, the positions and directions of tetrahedra
are usually adapted to vertices and edges of the graph underlying a
state to be acted on.  The coordinate volumes of tetrahedra then
replace the integration measure: $\epsilon^{abc}\md^3x\to
\delta^3\epsilon^{IJK}s_I^as_J^bs_K^c$. Moreover, internal tensors can
be written in terms of Pauli matrices, such as $\epsilon_{kmn}= -4{\rm
tr}(\tau_k\tau_m\tau_n)$. The tangents $s_I^a$, factors of $\delta$
and Pauli matrices can then be combined with Poisson brackets to obtain
\begin{equation}
 \tau_k\delta s_I^a\{{\cal A}_a^k,O\} \to -\frac{1}{i\hbar} h_{s_I}
 [h_{s_I}^{-1},\hat{O}]
\end{equation}
in terms of holonomies with their well-defined quantization, where $O$
could be the volume if (\ref{identity1}) is used, or the integrated
trace of extrinsic curvature $K$. For fine triangulations, $\delta\ll
1$, the error in replacing connection components by holonomies is
small, and it goes to zero in the limit where all edge lengths of
tetrahedra vanish. Similarly, covariant derivatives can be combined to
$\delta s_I^a{\cal D}_a$ and then regularized to a difference of
values at the endpoints of a small edge in direction $s_I^a$. If there
are always three factors where $\delta$ can be absorbed and the
quantized contributions vanish only when acting on vertices of a
graph, a well-defined operator results even in the limit when the
regulator is removed because for finite graphs finitely many terms
remain in the triangulation sum.

We first turn to the matter terms which arise in
(\ref{hdhamiltonianconstraint}). Some of them agree with the Dirac
Hamiltonian used in \cite{QSDV}, and can thus be quantized along the
same lines. However, our analysis has provided extra terms which must
be ensured to have well-defined quantum expressions, too. The current
interaction terms can directly be quantized with fermion operators
and using
\begin{eqnarray*}
 \frac{{\rm sgn}\det(e_a^i)}{\sqrt{q}}&=&
\frac{1}{6q}\epsilon^{abc}\epsilon_{ijk} e^i_ae^j_be^k_c\\ &=&
\frac{36\epsilon^{abc}\epsilon_{ijk}}{\gamma^3\kappa^3} \{{\cal
A}_a^i,V^{1/3}\} \{{\cal A}_b^j,V^{1/3}\} \{{\cal A}_c^k,V^{1/3}\}
\end{eqnarray*}
for a quantizable expression in terms of commutators of holonomies and
the volume operator. Edge tangents of the holonomies for the three
Poisson brackets provide the elementary coordinate volumes of the
triangulation, while half-densitized fermions in the current products
will simply be vertex-wise operators.

Terms of the form $q^{-1/2}P^a_i {\cal D}_aO$ where $O$ is an
expression of fermions can be reformulated using $\gamma\kappa P^a_i=
\frac{{\rm sgn}\det(e_a^i)}{2}\epsilon^{abc}\epsilon_{ijk}e_b^je_c^k$ in which we
can again absorb the inverse $\sqrt{q}$ after expressing the co-triads
as Poisson brackets. Here, we will have two holonomies requiring an
edge tangent vector as well as the covariant derivative which will
become a directional derivative once the triangulation volumes are
expressed via edge vectors: we use the expansion
$h_e(\delta)O(e(\delta))-O(e(0))\approx \delta\dot{e}^a{\cal D}_a O$
where $h_e(\delta)$ is a holonomy along an edge $e$ of coordinate
length $\delta$. Also these terms can thus be quantized by standard
techniques, which involves a discretization of the derivative.

Finally, we have to turn $q^{-1}\epsilon^{ijk} P^a_i e_b^k {\cal D}_a
P^b_j$ into an expression which can be quantized. We first
rewrite this as
\begin{eqnarray*}
&& \frac{\gamma^2\kappa^2}{q}\epsilon^{ijk} P^a_i e_b^k \partial_a
P^b_j = -\frac{\gamma^2\kappa^2}{q}\epsilon^{ijk} P^a_i P^b_j 
 \partial_a e_b^k\\
 &=& -\frac{{\rm sgn}\det(e_d^i)}{\sqrt{q}} \epsilon^{abc} e^k_c 
\partial_a e_b^k
= -{\rm sgn}\det(e_d^i)\epsilon^{abc} \frac{e^k_c}{q^{1/4}} \partial_a
 \frac{e^k_b}{q^{1/4}}
\end{eqnarray*}
which provides two factors of co-triads and one partial
derivative. Each of them will be combined with a tangent vector to
provide either holonomies or a discretized derivative. The inverse
powers of $q^{1/4}$ can be absorbed by choosing appropriate positive
powers of volume in Poisson brackets expressing the co-triads. (Note
that this is the reason why we had to move one $q^{-1/4}$ past the
partial derivative, because absorbing a single $q^{-1/2}$ would
require the ill-defined logarithm of volume.)

For the gravitational part of the constraint, the curvature components
${\cal F}_{ab}^k$ appear in a term which can be expressed as
$\int\md^3x \epsilon^{abc}F_{ab}^k\epsilon_{cde} \epsilon^{ijk}
P_j^dP^e_k/\sqrt{|\det P|}$. After triangulation, this takes the form
$\epsilon^{IJK}s_I^as_J^bs_K^c {\rm tr}(F_{ab}^k\tau_k
\tau_l\{A_c^l,V\})$ which can be written in terms of holonomies via
$\epsilon^{IJK} {\rm tr} (h_{IJ}h_K\{h_K^{-1},V\})$.

It remains to quantize the extrinsic curvature terms, where our goal
is to express $K_a^i$ in terms of Poisson brackets such as
$\left\{{\cal A}_{a}^{i},K\right\}$ and $\left\{{\cal
A}_{a}^{i},V\right\}$ which can be promoted to commutators of
well-defined operators. In the torsion-free case the integrated
extrinsic curvature is used in the expression $K_{a}^{i} =
\frac{1}{\kappa \gamma} \left\{A_{a}^{i},K\right\}$ for extrinsic
curvature components. This relation, proven e.g.\ in \cite{Reality},
turns out to be one of the main places where torsion changes the
quantization procedure of the Hamiltonian constraint.  Viewing
(\ref{K}) as a functional of the canonical pair $({\cal
A}_a^i,P^b_j)$, i.e.\ expressing $K_a^i$ in terms of ${\cal A}_a^i$
and $\Gamma_a^i$, yields
\begin{widetext}
\[
\{{\cal A}_b^j(y), K\} =\kappa ({\cal A}_{b}^{j}(y) - 
\Gamma_{b}^{j}(y)) - \kappa \int_{\Sigma} \md^{3}x \ P^{a}_{i}(x)
\frac{\delta \Gamma^{i}_{a}(x)}{\delta P^{b}_{j}(y)}
=\kappa \gamma K_{b}^{j}(y)+ \frac{\kappa^{2}\gamma^{2}\theta}{4(1+
\gamma^{2})}
\left(\frac{1}{\gamma} \epsilon^{j}_{\ \ kl}e_{b}^{k}(y)J^{l}(y)+
\frac{1}{2}e_{b}^{j}(y)J^{0}(y)\right)\,.
\]
\end{widetext}
Here, we have used $\Gamma^{i}_{a}= \widetilde{\Gamma}^{i}_{a}+{\cal
C}^{i}_{a}$ (which only requires solutions to second class
constraints) in the second step together with (\ref{cbk1}) and the
fact that $\kappa \int_{\Sigma} \md^{3}x \ P^{a}_{i}(x)\frac{\delta
\widetilde{\Gamma}^{i}_{a}(x)}{\delta P^{b}_{j}(y)}=0$, which can be
proven by a direct calculation or using the fact that $\tilde{F} :=
\kappa \gamma \int_{\Sigma} \md^{3}x \ P^{a}_{i}(x)
\widetilde{\Gamma}^{i}_{a}(x)$ is the generating functional of
$\widetilde{\Gamma}$. (Due to the presence of torsion, unless
$\theta=0$ the functional $F:=\kappa \gamma \int_{\Sigma} \md^{3}x \
P^{a}_{i}(x)\Gamma^{i}_{a}(x)$ no longer generates a canonical
transformation to $(K_a^i,P_b^j)$ since $\left\{{\cal
A}_{a}^{i},F\right\}\neq \Gamma_{a}^{i}$. Many of the differences
between torsion and torsion-free canonical gravity are reflected in
this property of the canonical structure.)

Together with (\ref{identity1}) it is then straightforward to show that
\begin{widetext}
\begin{eqnarray}
\label{kai}
K_{a}^{i}
&=& \frac{1}{\gamma \kappa} \left\{{\cal A}_{a}^{i},K\right\}-
\frac{\theta}{2\gamma(1+\gamma^{2})\sqrt{q}} \ \epsilon^{i}_{\ \ kl}
\left\{{\cal A}_{a}^{k},V\right\}\sqrt{q}J^{l}-\frac{\theta}{4(1+\gamma^{2})\sqrt{q}}
\left\{{\cal A}_{a}^{i},V\right\}\sqrt{q}J^{0}\,.
\end{eqnarray}
With these classical identities, the contributions $H^{E}[N]$
and ${\cal T}[N]$ to the Hamiltonian constraint become
\begin{eqnarray}
\label{cn}
H^{E}[N]= \frac{1}{\kappa^{2}\gamma^{\frac{3}{2}}}\int_{\Sigma} 
\md^{3}x \ N(x) \epsilon^{abc}{\cal F}^k_{ab}(x)
\left\{A^k_{c}(x),V\right\} {\rm sgn}\det(e_d^l)\,,
\end{eqnarray}
and
\begin{eqnarray}
\label{tn}
{\cal T}[N]&=& \frac{1}{2\kappa^{2}\gamma}\int_{\Sigma} \md^{3}x \ N(x) 
\epsilon^{abc}\epsilon_{kmn}\left\{{\cal A}^k_{a}(x),V\right\}K_{b}^{m}K_{c}^{n} {\rm sgn}\det(e_d^l)
\nonumber\ \\
&=& \frac{1}{2\kappa^{4}\gamma^{3}}\int_{\Sigma} \md^{3}x \ N(x) 
\epsilon^{abc}\epsilon_{kmn}\left\{{\cal A}^k_{a}(x),K\right\}
\left\{{\cal A}^m_{b}(x),K\right\}\left\{{\cal A}^n_{c}(x),V\right\} {\rm sgn}\det(e_d^l)\\ 
&&-\frac{2\theta}{\gamma^{3}\kappa^{3}(1+\gamma^{2})}\int_{\Sigma} \md^{3}x \ 
N(x)\epsilon^{abc} \ \epsilon_{kmn} \ \epsilon^{n}_{\ \ ij}
\left\{{\cal A}_{a}^{k}(x),V^{\frac{1}{2}}\right\}
\left\{{\cal A}_{b}^{m}(x),K\right\}\left\{{\cal A}_{c}^{i}(x),V^{\frac{1}{2}}\right\}
\sqrt{q}J^{j}  {\rm sgn}\det(e_d^l)\nonumber\ \\
&&-\frac{\theta}{\gamma^{3}\kappa^{2}(1+\gamma^{2})}\int_{\Sigma} \md^{3}x \ 
N(x)\epsilon^{abc} \ \epsilon_{kmn} \left\{{\cal A}_{a}^{k}(x),
V^{\frac{1}{2}}\right\}\left\{{\cal A}_{b}^{m}(x),K\right\}\left\{{\cal A}_{c}^{n}(x),
V^{\frac{1}{2}}\right\}\sqrt{q}J^{0} {\rm sgn}\det(e_d^l) \nonumber\ \\
&&+\frac{27\theta^{2}}{8\gamma^{2}\kappa^{2}(1+\gamma^{2})^{2}}\int_{\Sigma} \md^{3}x \ 
N(x) \ \epsilon^{abc} \ \epsilon_{kmn} \ 
\epsilon^{m}_{\ \ ij}\left\{{\cal A}_{a}^{k}(x),V^{\frac{1}{3}}\right\}
\left\{{\cal A}_{b}^{i}(x),V^{\frac{1}{3}}\right\}\left\{{\cal A}_{c}^{n}(x),
V^{\frac{1}{3}}\right\}
\sqrt{q}J^{j} \sqrt{q}J^{0} {\rm sgn}\det(e_d^l) \nonumber\ \\
&&+\frac{27\theta^{2}}{32\gamma \kappa^{2}(1+\gamma^{2})^{2}}\int_{\Sigma} \md^{3}x \ 
N(x) \ \epsilon^{abc} \ \epsilon_{kmn} 
\left\{{\cal A}_{a}^{k}(x),V^{\frac{1}{3}}\right\}\left\{{\cal A}_{b}^{m}(x),
V^{\frac{1}{3}}\right\}\left\{{\cal A}_{c}^{n}(x),V^{\frac{1}{3}}\right\}
\sqrt{q}J^{0} 
\sqrt{q}J^{0} {\rm sgn}\det(e_d^l) \nonumber\ \\
&&+\frac{27\theta^{2}}{8\gamma^{3} \kappa^{2}(1+\gamma^{2})^{2}}\int_{\Sigma} \md^{3}x \ 
N(x) \ \epsilon^{abc} \ \epsilon_{mkn} 
\left\{{\cal A}_{a}^{j}(x),V^{\frac{1}{3}}\right\}\left\{{\cal A}_{b}^{m}(x),
V^{\frac{1}{3}}\right\}\left\{{\cal A}_{c}^{k}(x),V^{\frac{1}{3}}\right\}
\sqrt{q}J^{n} 
\sqrt{q}J_{j} {\rm sgn}\det(e_d^l) \,. \nonumber
\end{eqnarray}
\end{widetext}
Here, we have already absorbed inverse powers of $\sqrt{q}$ in the
Poisson brackets, while keeping one factor of $\sqrt{q}$ with each
current component to make the product quadratic in half-densities of
fermions without other metric components.

It is thus clear that the presence of torsion introduces non-trivial
additional terms in the gravitational Hamiltonian constraint when it
is written in a form suitable for quantization. 

While no changes to the torsion-free construction of the Hamiltonian
constraint are required for expressing ${\cal F}_{ab}^i$ and ${\cal
A}_a^i$ in terms of holonomies, there is a further difference to the
treatment of $K$ in \cite{QSDI}. This quantity is not directly related
to a basic variable, but can be obtained from a Poisson bracket
$\{H^E[1],V\}$ where both ingredients are already written as
quantizable functions of basic quantities.  With $\Gamma^{i}_{a}$
having contributions from torsion, we obtain, using
(\ref{traceofgamma}) and the trace of (\ref{cbk1}),
\begin{widetext}
\begin{eqnarray}
\label{k4}
\left\{H^{E}[1],V\right\}&=& \sqrt{\gamma}
\frac{\gamma \kappa}{2}\int_{\Sigma} \md^{3}x \ \left(\epsilon^{ij}_{\ \ k} 
P^{a}_{i}e^{c}_{j}\partial_{a}e_{c}^{k}+2P^{b}_{n}(\Gamma^{n}_{b}+
\gamma K^{n}_{b})\right)\nonumber\ \\
&=& \gamma^{3/2} \kappa\int_{\Sigma} \md^{3}x \ 
\left(P^{b}_{n}{\cal C}^{n}_{b}+\gamma P^{b}_{n}K^{n}_{b}\right)= 
\gamma^{\frac{3}{2}}K-\frac{3\theta}{4}\gamma^{\frac{3}{2}}
\frac{\gamma}{1+\gamma^{2}}\int_{\Sigma} \md^{3}x \ \sqrt{q}J^{0},
\end{eqnarray}
which implies
\begin{equation}
\label{k5}
K = \gamma^{-\frac{3}{2}}\left\{H^{E}[1],V\right\}-
 i\frac{6\alpha^{2}\gamma\kappa\theta}{(1+\gamma^{2})(1+\alpha^{2})}
\int_{\Sigma} \md^{3}x \
(\theta_{R}\pi_{\xi}\xi-\theta_{L}\pi_{\chi}\chi)\,.
\end{equation}
\end{widetext}
Again, the presence of torsion implies that $K$ can no longer be
expressed just as the Poisson bracket of
$H^{E}[1]$ and $V$; the extra
term involving the fermion charge density in (\ref{k5}) is necessary
if the torsion is included in the connection. This result is
consistent since splitting the torsion contribution from $K_{a}^{i}$
and taking the trace of (\ref{kam}) reduces $K$ to the Poisson bracket
$\gamma^{-\frac{3}{2}}\left\{H^{E}[1],V\right\}$ without any
extra terms.  The additional term in (\ref{k5}), however, does not
have much effect since it only depends on the canonical fermion
half-densities, and thus drops out of the Poisson bracket with ${\cal
A}_a^i$ in (\ref{tn}) which is the only form in which $K$ appears.

It is interesting to note that, for $\alpha = \gamma$, the equations
(\ref{kai}), (\ref{tn}), (\ref{k4}), and (\ref{k5}) take the standard
forms of the torsion-free case (without any extra terms) since
$\theta$ vanishes. This results since the torsion contribution to the
spatial spin connection, ${\cal C}_{a}^{i}$, vanishes for $\alpha =
\gamma$ when the fermion fields are expressed in half-densities as
shown in (\ref{correctedc}). Therefore, except for the extra terms in
(\ref{hdhamiltonianconstraint}), the strategy for a loop quantization
of the gravitational sector of gravity non-minimally coupled to
fermions is exactly the same as that in vacuum for $\alpha = \gamma$.
Although this is the case which was also addressed in
\cite{FermionAshtekar}, we emphasize that the complete canonical
derivation for real variables has to be done to recognize the roles of
all possible contributions to the variables and constraints. In
particular, there are extra terms in (\ref{hdhamiltonianconstraint})
whose correct form must be used to quantize the Hamiltonian
constraint.

For $\alpha\not=\gamma$, the quantization of the scalar constraint of
gravity with fermions demands the quantization of the non-trivial
extra terms in (\ref{tn}) in addition to the terms appearing in
(\ref{hdhamiltonianconstraint}). This can be carried out using the
standard strategy: All extra terms have the structure $\int\md^3x
N\epsilon^{abc}\epsilon_{kmn} \{{\cal A}_a^k,O_1\} \{{\cal
A}_b^m,O_2\} \{{\cal A}_c^i,O_3\} O^n_i$ where $O_1$, $O_2$ and $O_3$
are either powers of $V$ or $K$, and $O^n_i$ is
$\epsilon^n_{ij}\sqrt{q}J^j$, $\delta^n_i\sqrt{q}J^0$,
$\epsilon^n_{ij}qJ^0J^j$, $\delta^n_iq (J^0)^2$ and $qJ^nJ_i$,
respectively, in all the required terms.  The operators $\hat{O}_i$
are obtained either as the volume operator or its commutator with the
Euclidean part of the Hamiltonian constraint. The current terms also
provide vertex operators directly in terms of the smeared fermion
operators $\hat{\Xi}_v$ and $\hat{X}_v$. For $J_0$, this can directly
be multiplied with the commutators, while $J^i$ can be inserted into
the trace through $\tau_iJ^i$. We do not list the long expressions for
complete operators here, but it is clear now that well-defined
quantizations exist for all the extra terms. This provides
quantizations of all terms in (\ref{tn}), completing the quantization
of the gravitational constraint in the presence of torsion.

\subsection{Parity}

In loop quantum gravity, the parity behavior is not manifest because
the Ashtekar connection transforms as $\Gamma_a^i+\gamma K_a^i\mapsto
\Gamma_a^i-\gamma K_a^i$ under parity, which does not result in a
straightforward transformation of its holonomies. For states in the
connection representation, there is thus no simple parity
transformation on the Hilbert space for which one could check
invariance of the theory. Sometimes the relation between $K_a^i$ and
extrinsic curvature is changed in the definition of basic variables,
making use of ${\rm sgn}\det (e_c^j) K_{ab}e^b_i$ with a sign factor
which would make the redefined $K_a^i$ and thus the whole Ashtekar
connection invariant under a reversal of the triad
orientation. However, the symplectic structure would be invariant
under this transformation only if a corresponding sign factor is
included in the momentum, now being $\det (e_b^j) P^a_i$ instead of
$P^a_i$. This momentum would also be invariant under orientation
reversal. With all the basic gravitational variables being invariant
under orientation reversal, one would simply loose any possibility to
implement non-trivial parity transformations at all. Thus, the only
possibility is to work with a theory whose parity behavior is rather
concealed.

While this may appear only as a technical problem in vacuum or with
non-fermionic matter, it becomes acute in the presence of fermions and
torsion. (Note that a second order formalism, where fermions would not
imply torsion contributions to the connection and thus allow a parity
behavior as in the vacuum theory, is unnatural for the connection
variables of the Ashtekar formulation as it underlies the loop
quantization.) As our classical discussion in Sec.~\ref{s:Parity}
showed, the precise behavior of the variables and constraints under
parity transformations is no longer obvious in the presence of
torsion. Even classically, the behavior is fully determined only
on-shell, making use not only of the constraints but also of some
equations of motion. While the classical solution space turns out to
be parity invariant for any $\alpha$, specific torsion contributions
to $\Gamma_a^i$ and $K_a^i$ acquire a behavior different from the
torsion-free parity behavior unless $\alpha=\gamma$. This observation,
consistent with \cite{FermionAshtekar}, indicates that the situation
of parity after quantization, where information about solutions of
equations of motion cannot be used, may be much more involved.

In fact, now the non-trivial parity behavior is hidden in holonomies
used as basic operators. At the kinematical level, there is no way of
knowing what unitary transformation could possibly represent a change
in parity, given that even classically one would have to make use of
constraints and equations of motion to determine that. In the
classical case, the behavior of the theory under parity became obvious
only after explicitly splitting off the torsion contributions from the
basic variables --- a procedure which we are denied in the quantum
theory. Triads have a much simpler (and obvious) behavior under
parity, but this, too, is difficult to implement at the quantum level
because no triad representation exists in the full theory
\cite{NonCommFlux}. Thus, the triad transformation cannot simply be
implemented at the state level.

It is thus quite likely that loop quantum gravity provides for parity
violating effects especially once fermions are included, even if the
classical fermion interactions used preserve parity. With the hidden
nature of torsion contributions and parity in the quantum formulation,
the precise form and magnitude of those parity violating effects is
not easy to discern. But some implications can be explored either with
effective equations (in their canonical form as described in
\cite{EffAc,EffCons}) which would allow one to perform some of the
steps required in the classical analysis of parity, or with symmetry
reduced models. An advantage of the latter would be that
some models exist (such as those introduced in
\cite{IsoCosmo,HomCosmo,Spin,SphSymm,EinsteinRosenQuant}) which do
allow a triad representation and thus a more direct implementation of
parity transformations.

\section{Conclusions}

We have summarized results of a complete canonical formulation of
gravity non-minimally coupled to fermions in Ashtekar variables. This
includes generalizations of basic results in the recent and some older
literature, such as the torsion-mediated four-fermion interaction, and
puts them on a firm canonical basis. We have used this for a
demonstration of parity invariance of classical solutions, which
required us to derive all contributions to the Ashtekar connection
explicitly and to write several new versions of the canonical
constraints, with explicit or implicit torsion contributions. The
different forms of the constraints are needed to understand the parity
behavior, and they also facilitate comparisons with earlier
derivations and allow crucial cross-checks of the results. Here, we
have noticed that our analysis fills in several gaps of previously
available derivations and generalizes them to arbitrary non-minimal
coupling.

The main purpose of the paper, however, is to provide a better and
more complete foundation for the loop quantization of gravity coupled
to fermions than can be found in the existing literature. Also this
requires knowledge of the details given in the derivation of the
canonical formalism to appreciate which of the established
quantization steps of the torsion-free case go through in the presence
of torsion, and where adaptations may be necessary. Overall, we find
that the quantization of fermion fields and their dynamics given by
Thiemann and others goes through in a well-defined manner. In details,
however, we have clarified several steps where previously gaps
existed, although they were not always realized.  For all values of
the non-minimal coupling parameter $\alpha$ there are new terms in the
constraints due to torsion which are derived here in complete form.
We have shown that torsion contributions and terms which arise from
using half-densitized spinors cancel in the connection for the case
where the non-minimal coupling parameter $\alpha$ equals the
Barbero--Immirzi parameter $\gamma$. As a consequence, the presence of
fermions does not change the quantization procedure much in this case,
although there are still additional terms. For $\alpha\not=\gamma$, on
the other hand, several additional adaptations to the usual
construction steps of the Hamiltonian constraint operator are
necessary.

While our results do not challenge the previous claims that all fields
necessary for the standard model of particle physics can be quantized
by loop techniques, some of the details of a specific quantization
have to be corrected. As such Hamiltonians may become relevant for
phenomenological considerations, e.g.\ in cosmology
\cite{ImmirziLambda,FermionBBN}, a precise understanding of the
quantum states and dynamical operators is not only necessary for a
complete quantization but even for potential physical applications. In
particular, we have highlighted the fact that current constructions of
loop quantum gravity do not suffice to show that it exactly
preserves parity.

\begin{acknowledgments}
  We are grateful to Robert Scherrer for his questions about fermions
  in loop quantum gravity, which motivated this work. We thank Sergei
  Alexandrov, Ghanashyam Date and Simone Mercuri for discussions and
  for pointing out some relevant references. Initial stages of the
  work have benefited from discussions with Thomas Thiemann. This work
  was supported in part by NSF grant PHY0653127. Some of the work was
  done while MB visited the Erwin-Schr\"odinger-Institute, Vienna,
  during the program ``Poisson Sigma Models, Lie Algebroids,
  Deformations, and Higher Analogues,'' whose support is gratefully
  acknowledged.
\end{acknowledgments}

\begin{appendix}

\section{The su(2) Spin Connection $\Gamma^{i}_{a}$ on $\Sigma$}
\label{appen:Spinconnection}

\subsection{Torsion-free spin connection}

In the torsion-free case, an explicit expression for the su(2) valued
spin connection $\widetilde{\Gamma}^{i}_{a}$ can be derived from the
fact that the covariant derivative of a co-triad vanishes:
$D_{a}e^{i}_{b}=\partial_{a}e_{b}^{j}-\Gamma_{ab}^{c}e_{c}^{j}+
\widetilde{\Gamma}_{ai}^{\ \ j}e_{b}^{i}=0$. Thus,
$\widetilde{\Gamma}_{ak}^{\ \ j}=
-e_{k}^{b}(\partial_{a}e_{b}^{j}-\Gamma_{ab}^{c}e_{c}^{j})$ and
\begin{equation}
\label{covariantderivativeofspinconnection}
\widetilde{\Gamma}_{a}^{i} =\frac{1}{2} \epsilon^{ij}_{\ \
  k}\widetilde{\Gamma}_{aj}^{\ \ k}=-\frac{1}{2}\epsilon^{ij}_{\ \
  k}e^{b}_{j}(\partial_{a}e_{b}^{k}-\Gamma_{ab}^{c}e_{c}^{k}) 
\end{equation}
where $\Gamma_{ab}^{c}$ is the usual torsion-free Levi-Civita
connection for $\widetilde{\Gamma}_{aj}^{\ \ \
k}:=\widetilde{\Gamma}_{a}^{l}\epsilon_{jl}^{\ \ k}$ is used. 
With the definition of the Levi-Civita connection and
$q_{ab}:=e_{a}^{k}e_{b}^{k}$ we obtain
\begin{widetext}
\begin{eqnarray}
\label{spin1}
e_{c}^{j}\Gamma_{ab}^{c}=\frac{1}{2}\left(e^{jd}e_{b}^{k}
\partial_{a}e_{d}^{k}+2\partial_{(a}e_{b)}^{j}+e^{dj}e_{a}^{k}
\partial_{b}e_{d}^{k}-
e^{dj}e_{a}^{k}\partial_{d}e_{b}^{k}-e^{dj}e_{b}^{k}
\partial_{d}e_{a}^{k}\right)\,. 
\end{eqnarray}
Inserting (\ref{spin1}) into
(\ref{covariantderivativeofspinconnection}) , we finally obtain the
desired expression for the spin connection
\begin{eqnarray}
\label{gamma}
\widetilde{\Gamma}_{a}^{i} =-\frac{1}{2} \epsilon^{ij}_{\ \
  k}\widetilde{\Gamma}_{aj}^{\ \ \ k}&=& \frac{1}{2}\epsilon^{ij}_{\ \
  k}e^{b}_{j}(\partial_{a}e_{b}^{k}-\Gamma_{ab}^{c}e_{c}^{k}) =
\frac{1}{2}\epsilon^{ijk}e_{k}^{b}(2\partial_{[b}e_{a]}^{j}+
e^{c}_{j}e_{a}^{l}\partial_{b}e_{c}^{l})\,. 
\end{eqnarray}

The following expressions are useful for computing
$\Gamma^{i}_{a}$ with torsion from the variational equations in the
presence of fermions:
\begin{eqnarray}
\label{traceofgamma}
e^{a}_{i}\widetilde{\Gamma}^{i}_{a}=
-\frac{1}{2}\epsilon^{ijk}e^{b}_{k}e^{a}_{i}
\partial_{a}e_{b}^{j}=\frac{1}{2\sqrt{q}}
\epsilon^{abc}e_{c}^{j}\partial_{a}e_{b}^{j}\,, 
\end{eqnarray}
and
\begin{eqnarray}
\label{deltakl}
\delta^{k}_{l}\epsilon^{bcd}e_{c}^{n}\partial_{b}e_{dn}+
2\epsilon^{bcd}e_{d}^{k}\partial_{b}e_{c}^{l}&=&
{\rm sgn}\det(e_a^i)(\frac{\sqrt{q}}{2}\epsilon^{ijk}\epsilon_{ijl}
\epsilon^{mnp}e^{a}_{m}e^{b}_{n}
\partial_{b}e_{ap}+2\sqrt{q}\epsilon^{ijk}e^{a}_{j}e^{b}_{i}
\partial_{b}e_{a}^{l}) \nonumber\\
&=&
{\rm sgn}\det(e_a^i)\sqrt{q}\epsilon^{ijk}(2e^{a}_{i}e^{b}_{l}
\partial_{[a}e^{b}_{b]}+
e^{a}_{j}e^{b}_{i}\partial_{b}e_{a}^{l})\,. 
\end{eqnarray}
Finally, the Gauss constraint
$D_{b}P^{bm}=\partial_{b}P^{bm}+\epsilon_{ij}^{\ \ m}
\Gamma^{i}_{b}P^{bj}= \frac{1}{2(1+\gamma^{2})} \sqrt{q} J^{m}$ for
the densitized triad $P^a_i$ implies
\begin{eqnarray}
 \Gamma^{k}_{b}P^{bl}-\Gamma^{l}_{b}P^{bk}&=& -\epsilon_{m}^{\ \
   kl}\partial_{b}{P^{bm}}+\frac{1}{2(1+\gamma^{2})} \ \epsilon_{m}^{\
   \  kl}\sqrt{q} J^{m} \nonumber\ \\ 
&=&
\frac{{\rm
   sgn}\det(e_a^i)}{\gamma\kappa}\left(-\epsilon^{bcd}e_{d}^{l}
\partial_{b}
e_{c}^{k}+\epsilon^{bcd}e_{d}^{k}\partial_{b}e_{c}^{l}\right)+ 
\frac{1}{2(1+\gamma^{2})} \ \epsilon_{m}^{\ \  kl}\sqrt{q} J^{m}\,.
\end{eqnarray}

\subsection{Connection with torsion}
\label{appen:SpinconnectionTorsion}

Varying the action by connection components, we obtain
\begin{eqnarray}
\label{nmpartial-a}
\frac{\delta \cal L}{\delta {(\,^{-}\!\!A_{c}^{l})}}=
\frac{1+\gamma^{2}}{2} \epsilon^{j}_{\
  lk}P^{c}_{j}\omega_{t}^{\ k0}+
\frac{1+\gamma^{2}}{2} \epsilon^{j}_{\
  kl}P^{[c}_{i}N^{a]}({\,^{+}\!\!A}^{k}_{a}+{\,^{-}\!\!A}^{k}_{a})+
\frac{1+\gamma^{2}}{2\gamma \kappa}
\epsilon^{acd}\partial_{a}(e_{dl}N) \nonumber\ \\
 + \frac{\gamma^2(1+\gamma^{2})\kappa}{2\sqrt{q}}
 \frac{N}{\kappa}P^{a}_{[k}P^{c}_{l]}
 ({\,^{+}\!\!A}^{k}_{a}-{\,^{-}\!\!A}^{k}_{a})+ \frac{N^{c}}{4}\sqrt{q}
 \left(\gamma+\frac{1}{\alpha}\right)J_{l} - 
\frac{\gamma\kappa N}{4} P^{c}_{l}\left(\gamma+\frac{1}{\alpha}\right)
J^{0}
 -\frac{\gamma\kappa N}{4}\ \epsilon^{j}_{\ lk}P^{c}_{j}
\left(1-\frac{\gamma}{\alpha}\right)J^{k}=0\,,
\end{eqnarray}
which in the canonical formulation serves as one of the second class
constraints.  After expressing (\ref{nmpartial-a}) in terms of
$\Gamma^{i}_{a}$ and $K^{i}_{a}$ first and then contracting with
$e^{m}_{c}$, we obtain
\begin{eqnarray}
\label{partial-a1}
\frac{1+\gamma^{2}}{2\gamma \kappa} \epsilon^{m}_{\ \  lk}\sqrt{q} \
\omega_{t}^{\ k0}- \frac{1+\gamma^{2}}{2\gamma \kappa} \epsilon^{m}_{\
  \ kl}\sqrt{q} \ N^{a}K_{a}^{k}+\frac{1+\gamma^{2}}{2\gamma \kappa}
\sqrt{q} \ e^{a}_{i}e^{m}_{c}\epsilon^{i}_{\ kl}
N^{c}K_{a}^{k}+ {\rm sgn}\det(e_a^i)\frac{1+\gamma^{2}}{2\gamma \kappa}
\epsilon^{bcd}e^{m}_{c}e_{dl}\partial_{b}N \nonumber\ \\
+{\rm sgn}\det(e_a^i)\frac{1+\gamma^{2}}{2\gamma \kappa}
\epsilon^{bcd}e^{m}_{c}N\partial_{b}e_{dl} -
\frac{(1+\gamma^{2})}{2\gamma \kappa} \sqrt{q} \ N (e^{a}_{i}
\delta^{m}_{l}-e^{a}_{l} \delta^{m}_{i})\Gamma^{i}_{a}= -\frac{\beta
  N^{c}e^{m}_{c}}{4}\sqrt{q} J_{l} + \frac{N\theta}{4} \epsilon^{m}_{\ \
  lk}\sqrt{q}J^{k} + \frac{\beta N}{4}\delta^{m}_{l}
\sqrt{q}J^{0}\,.
\end{eqnarray}
Contracting it with $\delta^{l}_{m}$ and using
the Gauss constraint, this equation simplifies considerably to
\begin{eqnarray}
\label{partial-a2}
{\rm sgn}\det (e_a^i)\frac{1+\gamma^{2}}{2\gamma \kappa} N \epsilon^{bcd}
e^{l}_{c}\partial_{b}e_{dl}- (1+\gamma^{2}) N
P^{a}_{i}\Gamma^{i}_{a}=\frac{3}{4}\beta N \sqrt{q}
J^{0}\,.
\end{eqnarray}
Symmetrizing the indices $m$ and $l$ in (\ref{partial-a1}) and using
(\ref{partial-a2}) for $e^a_i\Gamma_a^i$, we obtain the following
symmetric combination of $P^{a}_{l}$ and $\Gamma^{m}_{a}$
\begin{eqnarray}
\label{partial-a3}
\gamma\kappa(P^{a}_{l}\Gamma^{m}_{a}+P^{a}_{m}\Gamma^{l}_{a})=
{\rm sgn}\det(e_a^i)(\delta^{m}_{l}\epsilon^{bcd}e^{n}_{c}\partial_{b}e_{dn}-
\epsilon^{bcd}e^{m}_{c}\partial_{b}e_{dl}-
\epsilon^{bcd}e_{cl}\partial_{b}e^{m}_{d})- 
\frac{\beta \gamma\kappa}{2(1+\gamma^{2})}\delta^{m}_{l}\sqrt{q}
J^{0}\,.
\end{eqnarray}
On the other hand, the second class constraints can be seen to provide
an equation $2\partial_bP^{bm}+ 2\epsilon_{i}{}^{jm} P^b_j\Gamma_b^i=
\theta\sqrt{q}J^m/(1+\gamma^2)$, or
\begin{eqnarray}
\label{commutatorgamma1}
\gamma\kappa(P^{al}\Gamma^{m}_{a}-P^{am}\Gamma^{l}_{a})=
{\rm sgn}\det(e_a^i)(\epsilon^{bcd}e^{m}_{d}\partial_{b}e_{c}^{l}+
\epsilon^{bcd}e_{cl}\partial_{b}e^{m}_{d})+\frac{\theta \gamma
  \kappa}{2(1+\gamma^{2})}\epsilon_{j}^{\ ml}\sqrt{q} J^{j}\,. 
\end{eqnarray}
Combining (\ref{partial-a3}) and (\ref{commutatorgamma1}) yields 
\begin{eqnarray}
\label{partial-a4}
2\gamma\kappa P^{al}\Gamma^{k}_{a}=
{\rm sgn}\det(e_a^i)(\delta^{kl}\epsilon^{bcd}e^{n}_{c}\partial_{b}e_{dn}+
2\epsilon^{bcd}e^{k}_{d}\partial_{b}e_{c}^{l})+
\frac{\gamma \kappa}{2(1+\gamma^{2})}
\left({\theta} 
  \ \epsilon_{j}^{\ kl}\sqrt{q}J^{j}- \beta \delta^{kl}\sqrt{q} J^{0}\right)\,.
\end{eqnarray}
Next, inserting (\ref{deltakl}) into (\ref{partial-a4}), we find 
\begin{eqnarray}
\label{partial-a5}
e^{c}_{l}\Gamma^{k}_{c}=\frac{1}{2}\epsilon^{ijk}
e^{a}_{i}(2e^{b}_{l}\partial_{[a}e^{j}_{b]}+e^{b}_{j}\partial_{a}e_{bl})  
+\frac{\gamma \kappa}{4(1+\gamma^{2})}\left({\theta} \
  \epsilon_{j}^{\ kl}J^{j}-\beta \delta^{kl}
  J^{0}\right)\,,
\end{eqnarray}
and finally (\ref{gammabk1}).
\end{widetext}

\end{appendix}


\begin{thebibliography}{10}

\bibitem{AshVar}
A. Ashtekar, Phys.\ Rev.\ D {\bf 36},  1587  (1987).

\bibitem{AshVarReell}
J.~F. Barbero~G., Phys.\ Rev.\ D {\bf 51},  5507  (1995).

\bibitem{Immirzi}
G. Immirzi, Class.\ Quantum Grav. {\bf 14},  L177  (1997).

\bibitem{HolstAction}
S. Holst, Phys.\ Rev.\ D {\bf 53},  5966  (1996).

\bibitem{FermionAshtekar}
S. Mercuri, Phys.\ Rev.\ D {\bf 73},  084016  (2006).

\bibitem{QSDI}
T. Thiemann, Class.\ Quantum Grav. {\bf 15},  839  (1998).

\bibitem{SugraAshtekar}
M. Tsuda, Phys.\ Rev.\ D {\bf 61},  024025  (2000).

\bibitem{SpinTorsion}
F.~W. Hehl, P. von~der Heyde, G.~D. Kerlick, and J.~M. Nester, Rev.\ Mod.\
  Phys. {\bf 48},  393  (1976).

\bibitem{FermionImmirzi}
A. Perez and C. Rovelli, Phys.\ Rev.\ D {\bf 73},  044013  (2006).

\bibitem{FermionTorsion}
L. Freidel, D. Minic, and T. Takeuchi, Phys.\ Rev.\ D {\bf 72},  104002
  (2005).

\bibitem{FermionImmirziNonMin} 
S.\ Alexandrov, arXiv:0802.1221.

\bibitem{FermionHiggs}
T. Thiemann, Class.\ Quantum Grav. {\bf 15},  1487  (1998).

\bibitem{QSDV}
T. Thiemann, Class.\ Quantum Grav. {\bf 15},  1281  (1998).

\bibitem{LoopFermion1}
H.~A. Morales-T\'ecotl and C. Rovelli, Phys.\ Rev.\ Lett. {\bf 72},  3642
  (1994).

\bibitem{LoopFermion2}
H.~A. Morales-T\'ecotl and G. Esposito, Nuovo Cim.\ B {\bf 109},  973  (1994).

\bibitem{LoopFermion3}
H.~A. Morales-T\'ecotl and C. Rovelli, Nucl.\ Phys.\ B {\bf 451},  325  (1995).

\bibitem{NPZRev}
H. Nicolai, K. Peeters, and M. Zamaklar, Class.\ Quantum Grav. {\bf 22},  R193
  (2005).

\bibitem{ALMMT}
A. Ashtekar {\it et~al.}, J.\ Math.\ Phys. {\bf 36},  6456  (1995).

\bibitem{AreaVol}
C. Rovelli and L. Smolin, Nucl.\ Phys.\ B {\bf 442},  593  (1995), erratum:
  Nucl.\ Phys.\ B, {\bf 456}, 753 (1995).

\bibitem{Area}
A. Ashtekar and J. Lewandowski, Class.\ Quantum Grav. {\bf 14},  A55  (1997).

\bibitem{Vol2}
A. Ashtekar and J. Lewandowski, Adv.\ Theor.\ Math.\ Phys. {\bf 1},  388
  (1997).

\bibitem{RS:Spinnet}
C. Rovelli and L. Smolin, Phys.\ Rev.\ D {\bf 52},  5743  (1995).

\bibitem{SpinNet}
J.~C. Baez,  in {\em The Interface of Knots and Physics}, edited by L. Kauffman
  (American Mathematical Society, Providence, 1996), pp.\ 167--203.

\bibitem{Reality}
T. Thiemann, Class.\ Quantum Grav. {\bf 13},  1383  (1996).

\bibitem{NonCommFlux}
A. Ashtekar, A. Corichi, and J. Zapata, Class.\ Quantum Grav. {\bf 15},  2955
  (1998).

\bibitem{EffAc}
M. Bojowald and A. Skirzewski, Rev.\ Math.\ Phys. {\bf 18},
713--745 (2006).

\bibitem{EffCons}
M. Bojowald, B. Sandh\"ofer, A. Skirzewski and A. Tsobanjan, arXiv:0804.3365.

\bibitem{IsoCosmo}
M. Bojowald, Class.\ Quantum Grav. {\bf 19},  2717  (2002).

\bibitem{HomCosmo}
M. Bojowald, Class.\ Quantum Grav. {\bf 20},  2595  (2003).

\bibitem{Spin}
M. Bojowald, G. Date, and K. Vandersloot, Class.\ Quantum Grav. {\bf 21},  1253
   (2004).

\bibitem{SphSymm}
M. Bojowald, Class.\ Quantum Grav. {\bf 21},  3733  (2004).

\bibitem{EinsteinRosenQuant}
K. Banerjee and G. Date, arXiv:0712.0687.

\bibitem{FermionBBN}
M. Bojowald, R. Das, and R. Scherrer, Phys.\ Rev.\ D {\bf 77}, 084003
(2008).

\bibitem{ImmirziLambda}
S. Alexander and D. Vaid, hep-th/0702064.

\bibitem{Wald}
R.~M. Wald, {\em General Relativity} (The University of Chicago Press, Chicago,
  1984).

\bibitem{ThomasRev}
T. Thiemann, {\em Introduction to Modern Canonical Quantum General Relativity} 
(Cambridge University Press, Cambridge, UK, 2007).

\bibitem{Rov}
C. Rovelli, {\em Quantum Gravity} (Cambridge University Press, Cambridge, UK,
  2004).

\end{thebibliography}
\end{document}